\documentclass[twocolumn]{aastex7}

\begin{document}

\title{Characterizing TESS-Identified Quadruple and Higher Order Eclipsing Binaries:
\\I. Speckle Imaging with DSSI and HRCam 
}

\author{Steven R. Majewski}
\affiliation{Department of Astronomy, University of Virginia,
530 McCormick Rd., Charlottesville, VA 22904, USA}
\email[show]{srm4n@virginia.edu}

\author[0009-0007-1284-7240]{James W. Davidson Jr.}
\affiliation{Department of Astronomy, University of Virginia,
530 McCormick Rd., Charlottesville, VA 22904, USA}
\email[show]{jimmy@virginia.edu}

\author[0000-0002-4235-6369]{Robert F. Wilson}
\affiliation{Department of Astronomy, University of Maryland, College Park, MD 20742, USA }
\affiliation{NASA Goddard Space Flight Center, Greenbelt, MD 20771, USA}
\email{robert.f.wilson@nasa.gov}

\author[0000-0003-2159-1463]{Elliott P. Horch}
\altaffiliation{Adjunct Astronomer, Lowell Observatory}
\affiliation{Department of Physics, Southern Connecticut State University, 501 Crescent Street, New Haven, CT 06515, USA}
\email{horche2@southernct.edu}

\author[0009-0002-4246-878X]{Paul M. McKee}
\affiliation{Department of Astronomy, University of Virginia,
530 McCormick Rd., Charlottesville, VA 22904, USA}
\email{pmm3w@virginia.edu}

\author[0000-0002-2776-9827]{Evan Fagan}
\affiliation{Department of Astronomy, University of Virginia,
530 McCormick Rd., Charlottesville, VA 22904, USA}
\email{ef4taq@virginia.edu}

\author[0009-0009-9083-4415]{Gabriel N. Hartwell}
\affiliation{Department of Astronomy, University of Virginia,
530 McCormick Rd., Charlottesville, VA 22904, USA}
\email{gnh3pgz@virginia.edu}

\author[0000-0001-9786-1031]{Veselin B. Kostov}
\affiliation{NASA Goddard Space Flight Center, 8800 Greenbelt Road, Greenbelt, MD 20771, USA}
\affiliation{SETI Institute, 189 Bernardo Ave, Suite 200, Mountain View, CA 94043, USA}
\email{veselin.b.kostov@nasa.gov}

\author{Cassiopeia N. Adams}
\affiliation{Department of Physics, Southern Connecticut State University, 501 Crescent Street, New Haven, CT 06515, USA}
\email{Casnova.42@yahoo.com}

\author[0009-0007-4277-0360]{Torrie Sutherland}
\altaffiliation{Current address: Apache Point Observatory, 2001 Apache Point Road, Sunspot, New Mexico, NM 88349, USA}
\affiliation{Department of Physics, Southern Connecticut State University, 501 Crescent Street, New Haven, CT 06515, USA}
\email{torrieps@nmsu.edu}

\author[0000-0002-2084-0782]{Andrei Tokovinin}
\affiliation{Cerro Tololo Inter-American Observatory,
Casilla 603,
La Serena, Chile}
\email{andrei.tokovinin@noirlab.edu}

\author[0000-0001-7828-7257]{John C. Wilson}
\affiliation{Department of Astronomy, University of Virginia,
530 McCormick Rd., Charlottesville, VA 22904, USA}
\email{jcw6z@virginia.edu}

\correspondingauthor{James W. Davidson Jr.}

\begin{abstract}
    \added{NASA's TESS mission has unveiled a plethora of eclipsing binaries (EBs), among them hundreds of triples and higher order, hierarchical systems.} These complex targets require follow-up observations to enable full characterization of system architectures and identify the most compact multiples expected to undergo the most dramatic dynamical evolution. We report first results from a long-term effort to perform such follow-up, focusing here on multi-band speckle imaging of a majority, 57, of the sample of 97 quadruple and higher order eclipsing binaries (Q+EBs) identified via TESS light curves by \cite{Kostov2022}. Diffraction-limited imaging with the Differential Speckle Survey Instrument (DSSI) on the ARC 3.5-meter telescope and HRCam on the SOAR 4.1-m telescope reveals nearly 60\% of the 57 to resolve into two sources separated by $\geq$ 0.03 arcseconds. For these partly resolved systems, we report derived characteristics (e.g., relative position angle, angular separation, and magnitude differences in multiple passbands) from the speckle imaging. We find those Q+EBs partly resolved with 4-m class telescopes to have significantly inflated Gaia parallax errors and large Gaia RUWE, particularly for systems with separations comparable to Gaia's resolution limit ($\sim$0.6 arcseconds). For unresolved systems we report upper limits on angular and linear projected separations. We find two partly resolved Q+EBs with wide linear separations having eclipse timing variations that are therefore candidates of higher than quadruple multiplicity. Finally, we demonstrate how speckle imaging of resolved Q+EBs during an eclipse can clarify which speckle-resolved Q+EB subsystem is associated with a particular set of TESS eclipses.
\end{abstract}


\section{Introduction} \label{sec:intro}

About 40\% of star systems contain at least two stars gravitationally bound to and orbiting each other \citep{raghavan2010,duchene2013}; however, that percentage is only $\sim$2\% for more complex systems of four companions \citep{Eggleton2008}. These percentages increase for higher stellar masses \citep{Moe2017}. Multi-star systems are more than a statistical curiosity, however, as their dynamics hold clues to their formation and subsequent evolution \citep{Powell2021}. Hierarchical systems, in which a larger grouping of stars is subdivided into constituent subgroups, usually binaries, are especially helpful to  mine for information on the physics of binary and multiple star system formation. Such work is needed before we can confidently discriminate between competing origin models, such as multiplicity arising from gravitational capture versus creation through protostellar cloud fragmentation \citep{Tobin2016,Kostov2022}. Apart from revealing the physics of their formation, binaries, and especially higher order systems, merit more rigorous examination because of their importance as precursors to numerous astrophysical phenomena, from gravitational wave sources to the Type Ia supernovae that are so critical as standard candles on cosmological distance scales \citep[e.g.,][]{Fang2018}. 

Because of the additional information that can be extracted from them, eclipsing binaries (EBs) are a particularly valuable resource. The $\sim$$90^{\circ}$ inclination of the EBs to the line-of-sight is, in large part, unambiguous, which allows the determination of not only orbital periods, but relative stellar radii and effective temperatures --- necessary features to characterize stellar system architectures fully \citep[e.g.,][]{Soderhjelm2005}. EBs showing multiple families of eclipses --- a rare subset of higher-order systems and EBs generally --- are therefore of extreme interest. One configuration for such hierarchical systems is a quadruple arrangement consisting of a pair of individual binaries, which are themselves bound to and orbit each other; the famous “double double” $\epsilon$ Lyrae is a classical archetype of such a system. However, ``3+1'' systems (see, e.g., Fig.~1 of \citealt{Tokovinin2021}; also \citealt{Powell2023}) are also possible, apparently making up approximately 20\% of quadruple systems \citep{Tokovinin2014}. 

NASA's TESS mission \citep{Ricker2015}, via its all-sky survey, has discovered thousands of eclipsing binaries \citep{Prsa2022}. 
\added{Among this collection, a very small fraction show evidence for higher order hierarchies, including triple, quadruple, quintuple, and even sextuple candidates \citep[examples include][]{Rappaport2022,Rappaport2023,Rappaport2024,Kostov2022,Kostov2024}.} An initial list of 97 TESS quadruple/sextuple eclipsing binaries (Q$^+$EBs) was found by both machine-learning methods as well as visual inspection \citep{Kostov2022}. The unique aspect of this catalog is that the identified systems are all single TESS sources that nevertheless reveal two sets of eclipses with different periods; thus they are candidates for being intrinsically compact hierarchical systems with the potential to show short period dynamical evolution induced by tidal or mass transfer interactions.  To search for such evolution, and even to unravel the organization of these complex systems, requires follow-up observations.  For example, new light curve data can be used to look for eclipse timing variations and, if carried out in multiple bandpasses, aid in the determination of the constituent stellar types via photometric modeling.  Such information can also be gleaned from the families of spectral lines and Doppler shifts expressed in time series spectroscopic observations.  

At a more fundamental level, the Q$^+$EB candidates, like any TESS discoveries, must be followed up with ground-based observations simply to identify contaminating false positives. To enable the efficient creation of a full-sky survey, each of the TESS pixels spans an enormous 21 arcseconds, increasing the possibility that interesting photometric signals may be contaminated by the light of unrelated sources in the same pixel and thereby generating significant risk of false-positives. For the candidate double eclipsing systems, TESS data alone are unable to differentiate the interesting case of bound pairs of eclipsing binaries from the possibility, however remote, of independent eclipsing binaries separated by significantly less than a single TESS pixel. By design, the \citeauthor{Kostov2022} Q$^+$EBs feature components that are tightly packed at TESS resolution, and most are unresolvable even at the conventional $\sim$1 arcsecond seeing limit of ground-based optical telescopes. Thus, additional insights into the Q$^+$EB architectures can be provided by higher resolution imaging,  whether or not the targets are resolved into constituent subsystems, since {\it lack} of system resolution into parts provides even more stringent limits on system compactness and points to systems of greater potential interest in terms of complex dynamical interaction and evolution.

Traditional ground-based imaging is blurred by the high frequency transient effects of Earth's atmospheric turbulence. However, speckle imaging presents a powerful and efficient means of overcoming that limitation and achieving diffraction-limited resolution that may begin to resolve components of these systems using ground-based telescopes.  Here we report on a pilot program to undertake diffraction-limited imaging of the \cite[][; K22 hereafter]{Kostov2022} Q$^+$EB sample using the Differential Speckle Survey Instrument (DSSI; \citealt{Horch2009}) on the Astrophysical Research Consortium (ARC) 3.5-meter telescope at Apache Point Observatory (APO) \citep{Davidson2024}.  We have observed all 64 of the APO-accessible sources above a declination of approximately $-32^{\circ}$ in the K22 catalog at least once, albeit not always in optimal observing conditions. In the end, we have successfully obtained results for 38 of the APO-accessible sources; those for which we have been unsuccessful in obtaining a reliable result generally are the fainter sources (e.g., $G \gtrsim 13$).
Our APO observations have been supplemented by observations of 23
Southern Hemisphere systems from K22 using the HRCam instrument on the SOAR 4.1-m telescope \citep{Tokovinin2008,Tokovinin2018}. Four systems have been observed with both instruments. Therefore the total number of systems observed from the K22 catalog is 57. A primary result of these observations described here is that $\sim$60\% (34 of the 57) of the observed Q$^+$EBs have been resolved into at least two subsystems, and for these the angular separation and position angle of the subcomponents are provided. For the Q$^+$EBs we have not resolved, new limits are placed on the compactness and/or contrast of the systems based on the null result (i.e., ``high quality non-detection of resolution into more than one source'').

More recently, and subsequent to the initiation of the speckle imaging program described here, \citet{Kostov2024} published a second list of 101 additional Q$^+$EBs found in TESS full frame images.  Because we have only just started to undertake follow-up work on these more recently identified systems, whereas we are much more complete in our observations of the initial K22 list of 97 systems, for the remainder of this paper we focus entirely on the candidates from that K22 list.

In Section \ref{sec:observations} of this paper we describe the speckle imaging observations.  Section \ref{sec:measured} gives the results of these observations, including a tabulation of the determined position angles, separations, and magnitude differences (typically in multiple passbands) for the partly resolved Q$^+$EBs, and an assessment of the resolvability limits for the unresolved systems for which we have high quality observations. In Section \ref{sec:distances} we assess the distances to the systems, which are needed to convert angular separations into linear separations for resolved systems and to provide estimates of the upper limit of linear size on the unresolved Q$^+$EBs.  We show that the {\it Gaia}-parallax-based distances typically have large uncertainties for the speckle resolved systems with separations below $\sim$$0.6$ arcsecond. However, reasonable parallax distances are possible for those systems that we do not resolve with high quality speckle data. With these distance estimates in hand we discuss several interesting systems --- e.g., those seemingly of extreme compactness --- suggested by our observations in Section \ref{sec:special}, as well as objects for which we have apparently resolved two subcomponents, but for which we believe that one of the subcomponents may not be part of the Q$^+$EB at all but may well represent yet another, even higher order hierarchy (e.g., quintuple) or a line-of-sight contaminant. Finally, in Section \ref{sec:SIDEs} we demonstrate how speckle imaging obtained during eclipses can be used to assign particular eclipses with specific speckle-imaging-resolved subsystems within a Q$^+$EB.

The present speckle imaging effort represents one facet of a multidimensional observational campaign we are undertaking to characterize the TESS-identified hierarchical EBs, and which also includes (1) new multiband transit photometry to look for timing transit variations, improve the system ephemerides, and give more constraints on the photometric building blocks of the systems, (2) high resolution time series spectroscopy to derive spectral types from multilined systems and to map Doppler shifts to constrain orbits, and (3) speckle imaging taken during eclipses, through which we can assign specific eclipse families to specific resolved source components as discussed in Section \ref{sec:SIDEs}.

\section{Observations}
\label{sec:observations}

Speckle imaging of the K22 sample was undertaken primarily with the DSSI instrument \citep{Horch2009}, recently commissioned on the ARC 3.5-m telescope at APO in New Mexico \citep{Davidson2024}.  These observations were obtained over the course of seven quarterly observing runs from 2022 Q2 to 2023 Q4, during which DSSI was block-scheduled and used to collect data for a variety of science programs, with targeting for different programs intermingled to make optimal use of observing time, conditions, and calibration observations. In some cases, specific Q$^+$EB targets were observed over multiple epochs, to (1) check the consistency of speckle imaging results from run to run, (2) confirm prior results taken under less favorable conditions, and (3) search for potential astrometric changes (as unlikely as those might be over the relatively short baseline). Regarding (1), \cite{Davidson2024} have shown that, despite repeated movement of DSSI on and off the telescope, as well as the movement of optics within DSSI as part of efforts to test new capabilities, the performance of DSSI from run to run is highly repeatable. Using observations during the first four observing runs (i.e., the first year of operation), \cite{Davidson2024} found that DSSI delivers an astrometric precision of $2.06 \pm 0.11$ mas and a photometric precision of $0.14 \pm 0.04$ magnitudes when mounted on the ARC 3.5-m telescope. Because the bulk of the DSSI data reported here were included in the analysis of the instrument performance in \citet{Davidson2024}, we adopt the above values as representative of the precision of all DSSI results presented here (although we revisit the astrometric reliability of our Q$^+$EB measurements via repeat observations below, in \ref{sec:astrometry_repeat} and \ref{sec:photometry_repeat}).

By use of a dichroic to split the beam, DSSI takes speckle observations simultaneously in two passbands; for the Q$^+$EBs, speckle images were taken in passbands with central wavelengths of 692 and 880nm with widths of 40 and 50nm, respectively. Single frame exposure times were 40ms. A minimum of 1000 exposures in each filter were obtained for each source, although for fainter sources or during non-ideal observing conditions, as many as 15,000 individual exposures were collected. 

\begin{figure*}
    \centering
    \includegraphics[width=0.195\textwidth]{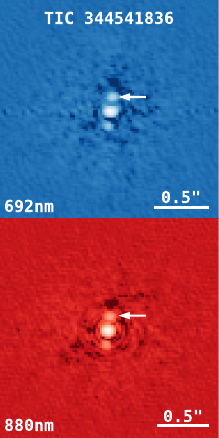}
    \includegraphics[width=0.195\textwidth]{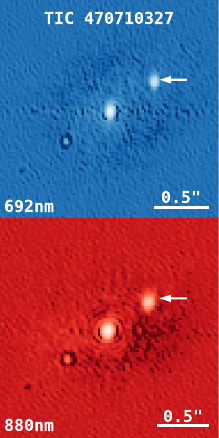}
    \includegraphics[width=0.195\textwidth]{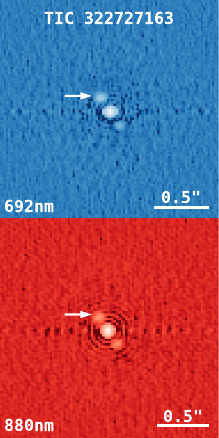}
    \includegraphics[width=0.195\textwidth]{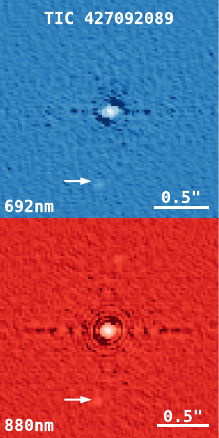}    
    \includegraphics[width=0.195\textwidth]{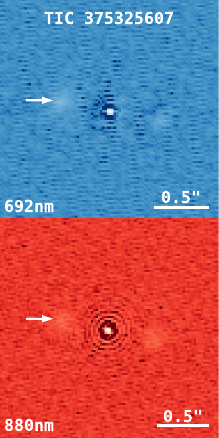}
    \caption{A representative sample of DSSI reconstructed images from the resolved systems in Table 1. Sources are organized left to right from brightest (TIC 344541836; $G=7.88$) to dimmest (TIC 375325607; $G=13.14$). The panels show the reconstructed image in the 692 nm channel ({\it top, blue-tinted panels}) and in the 880 nm channel ({\it bottom, red-tinted panels}). For each panel, North is up and East is to the left. The white arrow points to the real secondary source to distinguish it from the ghost image mirrored 180$^\circ$ in position angle. The flux scaling is logarithmic to highlight the detected sources. The $\Delta m$ contrasts between primary and secondary are given in Table \ref{tab:speckle measurements}.}
    \label{fig:ris}
\end{figure*}

The reduction of these data follows the ``standard''  procedures used for DSSI data taken at other telescopes, and variously described or summarized in \citet{Horch2011a}, \citet{Horch2021}, and \citet{Davidson2024}. Once the power spectrum and bispectrum of an observation are computed from the speckle data frames, the reduction method is essentially twofold: (1) a basic reconstructed image (e.g., Fig.~\ref{fig:ris}) is formed by assembling the diffraction-limited Fourier transform of the object from these spatial frequency spectra (including deconvolution by an appropriate point source observation and low-pass filtering to reduce noise), and (2) a weighted least-squares fit to fringes in the object's power spectrum is computed when a second source is identified in the reconstructed image. These fringe fits yield the separation, position angle, and magnitude difference of the secondary star relative to the primary.

Approximately one third of the K22 systems are not accessible from APO.  From this list, 23 were successfully observed using the HRCam speckle imager \citep{Tokovinin2008,Tokovinin2018} on the Southern Astrophysical Research (SOAR) 4.1-m telescope located at Cerro Pach\'on, Chile. The majority of these observations were taken at our request, however all of the HRCam observations presented here have been previously reported as part of the yearly, or bi-yearly, compilation of speckle observations from SOAR \citep{Tokovinin2021b,Tokovinin2022,Mason2023,Tokovinin2024}. The data reduction methodology for the HRCam observations is also independent, and is described in \citet{Tokovinin2010} and the corresponding data release papers.

\section{Results}
\label{sec:results}

\subsection{Measured Parameters}
\label{sec:measured}

A total of 64 Q$^+$EB targets from K22 were observed using DSSI on the ARC 3.5-m telescope and 23 targets have previously published observations with HRCam on the 4.1-m SOAR telescope, with four targets common to both. We report here only those observations with DSSI that have resulted in reasonably confident resolution of a Q$^+$EB into two subsystems, or for which we can cite a confident limit that a Q$^+$EB is unresolved --- i.e., a high-quality non-detection (HQND), that is to say an {\it unresolved} system. While the results from observations taken with HRCam can be found in \citet{Tokovinin2021b}, \citet{Tokovinin2022}, \citet{Mason2023}, and \citet{Tokovinin2024}, they are presented again here to tie them into the scientific context for which most of the observations were originally sought.
                    
Table \ref{tab:speckle measurements} summarizes the astrometric and photometric results for those targets we have partly resolved into two subsystems, with columns as follows: the TESS Input Catalog ID, the {\it Gaia} $G$ magnitude of the source, the UTC Date(s) the target was observed in decimal years, the derived position angle from the brighter to the fainter subsystem in the observed passband ($\theta$), their angular separation ($\rho$), the magnitude difference of the two components discerned ($\Delta m$), the wavelength of the observation ($\lambda$), the bandwidth of the filter ($\Delta \lambda$), and finally a discoverer designation for systems that have been previously resolved as double, as well as the reference for HRCam observations. We note that only eight of the 34 resolved targets are listed in the  Washington Double Star Catalog \citep{Mason2001} as having been previously resolved into more than one source, and all prior measurements are consistent with the speckle results reported here. For TIC 168789840, there are a number of previous speckle observations because this sextuple system is being monitored to search for a wobble in the AB-C positions caused by the AB subsystem \citep{Powell2021}. Observations with the $\Delta \lambda = 40$nm wide filter at 692nm and the $\Delta \lambda = 50$nm wide filter at 880nm were taken with DSSI on the ARC 3.5-m telescope, while those with the $\Delta \lambda = 170$nm wide filter at 824 nm and the $\Delta \lambda = 84$nm wide filter at 517nm were taken with HRCam on the SOAR 4.1-m telescope.  In no cases were more than two sources resolved among the Q$^+$EB targets. 

Figure \ref{fig:ris} gives examples of some of the reconstructed images in the simultaneously obtained 692nm and 880nm filters of Q$^+$EB sources that were resolved into two subcomponents using DSSI data for sources spanning from $G = 7.88$ to as faint as $G= 13.14$ and with secondary contrasts as large as $\Delta m \sim 3.8$ (for TIC 427092089 at $\lambda = 692$nm).  The figure shows clear resolution of binaries with separations as low as $\sim0.12"$ (though binaries with separations as small as $\sim0.06"$ were resolved with DSSI). The real position of the resolved companion is indicated with a white arrow, and is determined by computing the phase information for the data set with the image bispectrum methods \citep{Lohmann1983} using the relaxation technique of \citet{Meng1990} described in further detail in previous DSSI papers \citep[e.g.,][]{Horch2011a,Horch2021,Davidson2024}. Ghost images, mirrored 180$\circ$ in position angle, sometimes occur in our image reconstructions of fainter binaries, particularly if the two stars do not have a large magnitude difference. This is because we start with the so-called ‘zero phase’ assumption and iterate from that point to produce a phase map in the Fourier domain. Zero phase in the Fourier domain is equivalent to an image reconstruction that is the square-root of the autocorrelation, and thus has peaks of equal value on either side of the primary star. For binaries with smaller $\Delta m$ values and bispectra with low SNR, our phase reconstruction routine produces a phase map that only slowly changes as a function of the number of iterations performed and often does not converge to a smooth final image with only one peak for the secondary star. The resulting image can show two peaks, but with the true location of the secondary having a higher maximum value.

\startlongtable
\begin{deluxetable*}{lDDDDDccl}
\tabletypesize{\scriptsize}
\tablewidth{0pt} 
\tablecaption{Speckle Measurements of Resolved Systems\label{tab:speckle measurements}}
\tablehead{
\colhead{TIC \#} & \multicolumn2c{$G$} & \multicolumn2c{Date} & \multicolumn2c{$\theta$}& \multicolumn2c{$\rho$} & \multicolumn2c{$\Delta m$} & \colhead{$\lambda$} & \colhead{$\Delta\lambda$} & \colhead{Discov. \&} \\
\colhead{} & \multicolumn2c{(mag)} & \multicolumn2c{(2000+)} & \multicolumn2c{($^\circ$)} & \multicolumn2c{(")} & \multicolumn2c{(mag)} & \colhead{(nm)} & \colhead{(nm)} & \colhead{Reference\tablenotemark{a}}
} 
\decimals
\startdata 
25818450  & 11.45 & 22.74344 & 181.0 & 0.0633 & 1.51   & 692 & 40  &  \\
          & .     & 22.74344 & 181.2 & 0.0639 & 1.33   & 880 & 50  &  \\
45160946  & 13.70 & 23.9119  & 163.1 & 0.1869 & 1.8  : & 824 & 170 & HRCam1 \\
75740921  & 13.22 & 23.9119  & 95.9  & 0.3986 & 0.4  : & 824 & 170 & HRCam1 \\
78568780  & 11.51 & 22.74658 & 246.9 & 0.2632 & 1.26   & 692 & 40  &  \\
          & .     & 22.74658 & 247.0 & 0.2631 & 1.20   & 880 & 50  &  \\
89278612  & 11.12 & 22.74307 & 184.7 & 0.0894 & 0.61   & 692 & 40  &  \\
          & .     & 22.74307 & 186.2 & 0.0894 & 0.55   & 880 & 50  &  \\
97356407  & 6.31  & 22.86431 & 228.9 & 0.1103 & 4.33   & 692 & 40  &  \\
          & .     & 22.86431 & 231.9 & 0.1102 & 3.28   & 880 & 50  &  \\
          & .     & 23.3246  & 230.2 & 0.1050 & 3.4  : & 824 & 170 & HRCam1 \\
          & .     & 23.9117  & 236.0 & 0.1001 & 4.0  : & 824 & 170 & HRCam1 \\
146810480 & 8.96  & 20.2014  & 130.0 & 0.0882 & 0.9    & 824 & 170 & HDS1532, HRCam4 \\
          & .     & 22.1952  & 129.7 & 0.0992 & 0.9    & 824 & 170 & HRCam2 \\
161043618 & 12.27 & 22.35715 & 185.7 & 0.1074 & 0.31   & 692 & 40  &  \\
          & .     & 22.35715 & 184.8 & 0.1115 & 0.79   & 880 & 50  &  \\
168789840 & 11.67 & 20.8233  & 257.7 & 0.4230 & 0.3    & 824 & 170 & HRCam4 \\
          & .     & 20.8365  & 257.6 & 0.4233 & 0.3    & 824 & 170 & HRCam4 \\
          & .     & 20.9238  & 257.6 & 0.4235 & 0.3    & 824 & 170 & HRCam4 \\
          & .     & 20.9238  & 257.7 & 0.4243 & 0.3  : & 517 & 84  & HRCam4 \\
          & .     & 21.1616  & 258.1 & 0.4221 & 0.3    & 824 & 170 & HRCam3 \\
          & .     & 21.1616  & 258.1 & 0.4229 & 0.2  : & 517 & 84  & HRCam3 \\
          & .     & 22.1946  & 257.8 & 0.4228 & 0.4    & 824 & 170 & HRCam2 \\
          & .     & 22.73817 & 258.9 & 0.4190 & 0.82   & 692 & 40  &  \\
          & .     & 22.73817 & 257.9 & 0.4160 & 0.59   & 880 & 50  &  \\
          & .     & 22.7745  & 258.1 & 0.4231 & 0.2    & 824 & 170 & HRCam2 \\
          & .     & 23.5724  & 258.4 & 0.4234 & 0.3    & 824 & 170 & HRCam1 \\
200094011 & 9.71  & 22.73841 & 208.6 & 0.5439 & 1.10   & 692 & 40  & HDS 779 \\
          & .     & 22.73841 & 208.7 & 0.5438 & 1.10   & 880 & 50  &  \\
          & .     & 23.8980  & 208.4 & 0.5435 & 1.0    & 824 & 170 & HRCam1 \\
201310151 & 14.88 & 23.5720  & 138.4 & 0.1157 & 1.7  : & 824 & 170 & HRCam1 \\
239872462 & 11.29 & 22.74372 & 11.1  & 1.6365 & 1.35 * & 692 & 40  &  \\
          & .     & 22.74372 & 11.2  & 1.6385 & 0.61 * & 880 & 50  &  \\
255532033 & 10.74 & 23.6643  & 191.3 & 0.1124 & 0.8  : & 824 & 170 & HRCam1 \\
257776944 & 9.51  & 23.4889  & 229.3 & 0.1099 & 1.5    & 824 & 170 & HRCam1 \\
266771301 & 12.26 & 22.74110 & 215.2 & 0.5158 & 0.32   & 692 & 40  &  \\
          & .     & 22.74110 & 215.0 & 0.5168 & 0.79 * & 880 & 50  &  \\
          & .     & 22.87175 & 214.9 & 0.5177 & 0.00   & 692 & 40  &  \\
          & .     & 22.87175 & 216.0 & 0.5124 & 0.00   & 880 & 50  &  \\
283940788 & 11.73 & 22.74069 & 245.4 & 0.4890 & 3.76   & 692 & 40  &  \\
          & .     & 22.74069 & 245.1 & 0.4929 & 3.82   & 880 & 50  &  \\
286470992 & 10.11 & 22.74102 & 198.1 & 1.6336 & 1.82 * & 692 & 40  & MLR 108 \\
          & .     & 22.74102 & 198.1 & 1.6353 & 1.75 * & 880 & 50  &  \\
          & .     & 22.86632 & 198.1 & 1.6501 & 2.26 * & 692 & 40  &  \\
          & .     & 22.86632 & 198.0 & 1.6453 & 2.00 * & 880 & 50  &  \\
300446218 & 14.40 & 23.9118  & 99.0  & 0.0932 & 0.1  : & 824 & 170 & HRCam1 \\
307119043 & 10.01 & 22.74338 & 157.7 & 0.8281 & 2.05   & 692 & 40  & TDS 1706 \\
          & .     & 22.74338 & 157.7 & 0.8273 & 1.75   & 880 & 50  &  \\
317863971 & 10.32 & 22.87238 & 184.6 & 2.5691 & 3.81 * & 692 & 40  &  \\
          & .     & 22.87238 & 184.7 & 2.5669 & 3.23 * & 880 & 50  &  \\
          & .     & 23.18129 & 184.5 & 2.5793 & 3.92 * & 692 & 40  &  \\
          & .     & 23.18129 & 184.5 & 2.5830 & 3.36 * & 880 & 50  &  \\
322727163 & 11.00 & 22.74326 & 35.6  & 0.1621 & 1.79   & 692 & 40  &  \\
          & .     & 22.74326 & 36.2  & 0.1641 & 1.71   & 880 & 50  &  \\
328181241 & 11.14 & 23.66694 & 244.9 & 0.0644 & 0.70   & 692 & 40  &  \\
          & .     & 23.66694 & 249.2 & 0.0599 & 0.02   & 880 & 50  &  \\
336882813 & 11.87 & 22.74376 & 136.2 & 0.5943 & 1.88   & 692 & 40  &  \\
          & .     & 22.74376 & 136.2 & 0.5950 & 2.07   & 880 & 50  &  \\
          & .     & 23.18112 & 136.3 & 0.5993 & 1.82   & 692 & 40  &  \\
          & .     & 23.18112 & 136.6 & 0.5993 & 1.70   & 880 & 50  &  \\
344541836 & 7.88  & 22.36013 & 354.0 & 0.1229 & 1.36   & 692 & 40  & MLR 590 \\
          & .     & 22.36013 & 352.5 & 0.1209 & 0.78   & 880 & 50  &  \\
          & .     & 22.74327 & 352.8 & 0.1418 & 1.79   & 692 & 40  &  \\
          & .     & 22.74327 & 352.9 & 0.1403 & 1.68   & 880 & 50  &  \\
          & .     & 22.87168 & 352.8 & 0.1409 & 1.72   & 692 & 40  &  \\
          & .     & 22.87168 & 353.4 & 0.1395 & 1.62   & 880 & 50  &  \\
348651800 & 11.66 & 22.87266 & 56.5  & 0.4088 & 1.36   & 692 & 40  &  \\
          & .     & 22.87266 & 56.9  & 0.4125 & 1.02   & 880 & 50  &  \\
          & .     & 23.18132 & 56.8  & 0.4203 & 1.49   & 692 & 40  &  \\
          & .     & 23.18132 & 56.5  & 0.4228 & 1.26   & 880 & 50  &  \\
          & .     & 23.8983  & 56.9  & 0.4127 & 1.2    & 824 & 170 & HRCam1 \\
357810643 & 7.01  & 23.3248  & 108.3 & 0.4419 & 2.7    & 824 & 170 & HDS1336, HRCam1 \\
367448265 & 7.91  & 22.74369 & 10.5  & 0.6603 & 2.95   & 692 & 40  & HDS 691 \\
          & .     & 22.74369 & 10.5  & 0.6631 & 2.64   & 880 & 50  &  \\
          & .     & 23.18385 & 10.5  & 0.6728 & 2.99   & 692 & 40  &  \\
          & .     & 23.18385 & 10.5  & 0.6681 & 2.64   & 880 & 50  &  \\
          & .     & 23.67531 & 10.7  & 0.6557 & 3.02   & 692 & 40  &  \\
          & .     & 23.67531 & 10.6  & 0.6587 & 2.71   & 880 & 50  &  \\
375325607 & 13.14 & 22.87147 & 79.3  & 0.4847 & 0.67   & 692 & 40  &  \\
          & .     & 22.87147 & 79.5  & 0.4886 & 0.39   & 880 & 50  &  \\
389836747 & 10.78 & 22.74064 & 53.2  & 0.1528 & 0.92   & 692 & 40  &  \\
          & .     & 22.74064 & 53.3  & 0.1521 & 0.98   & 880 & 50  &  \\
          & .     & 22.86637 & 52.2  & 0.1548 & 0.90   & 692 & 40  &  \\
          & .     & 22.86637 & 53.1  & 0.1547 & 0.82   & 880 & 50  &  \\
399492452 & 10.83 & 23.4124  & 333.6 & 0.0325 & 0.0  : & 824 & 170 & HRCam1 \\
427092089 & 12.27 & 22.74332 & 172.4 & 0.7165 & 3.77   & 692 & 40  &  \\
          & .     & 22.74332 & 172.2 & 0.7126 & 3.46   & 880 & 50  &  \\
438226195 & 13.15 & 22.87215 & 290.8 & 0.4601 & 0.98   & 692 & 40  &  \\
          & .     & 22.87215 & 292.1 & 0.4579 & 0.61   & 880 & 50  &  \\
461614217 & 10.44 & 23.3246  & 107.5 & 0.0644 & 1.0  : & 824 & 170 & HRCam1 \\
470710327 & 9.79  & 22.74342 & 304.9 & 0.5112 & 1.19   & 692 & 40  & MLR 24AB \\
          & .     & 22.74342 & 304.8 & 0.5110 & 1.13   & 880 & 50  &  \\
          & .     & 23.79756 & 305.2 & 0.5103 & 1.15   & 692 & 40  &  \\
          & .     & 23.79756 & 305.3 & 0.5069 & 1.12   & 880 & 50  &  \\
          & .     & 23.79778 & 305.2 & 0.5096 & 1.18 * & 692 & 40  &  \\
          & .     & 23.79778 & 305.2 & 0.5078 & 1.11 * & 880 & 50  &  \\
          & .     & 23.79800 & 305.2 & 0.5065 & 1.23   & 692 & 40  &  \\
          & .     & 23.79800 & 305.2 & 0.5045 & 1.16   & 880 & 50  &  \\
          & .     & 23.80023 & 305.2 & 0.5079 & 1.26   & 692 & 40  &  \\
          & .     & 23.80023 & 305.2 & 0.5073 & 1.13   & 880 & 50  &  \\
          & .     & 23.80061 & 305.2 & 0.5082 & 1.08   & 692 & 40  &  \\
          & .     & 23.80061 & 305.1 & 0.5074 & 1.02   & 880 & 50  &  \\
          & .     & 23.80078 & 305.2 & 0.5068 & 1.09   & 692 & 40  &  \\
          & .     & 23.80078 & 305.1 & 0.5065 & 1.07   & 880 & 50  &  \\
          & .     & 23.80100 & 305.2 & 0.5064 & 1.22   & 692 & 40  &  \\
          & .     & 23.80100 & 305.2 & 0.5069 & 1.12   & 880 & 50  &  \\
\enddata
\tablecomments{\added{For all DSSI observations, we adopt the astrometric precision of 2.06$\pm$0.11 mas and a photometric precision of 0.14$\pm$0.04 magnitudes \citep{Davidson2024}. For all HRCam observations, the average astrometric precision is $\sim$1-2 mas and the average photometric precision is $\sim$0.1-0.2 magnitudes \citep{Tokovinin2022}, similar to DSSI.} The $\Delta m$ values marked with an asterisk (*) are observations from DSSI that have a (seeing)$\times$(separation) $>$ 0.6 arcsec$^2$. These are considered an upper limit because the observation may be affected by speckle decorrelation as discussed in \citet[][see also Section \ref{sec:photometry_repeat}]{Horch2004}. Similarly, $\Delta m$ values marked with a colon (:) are observations from HRcam that have a signal-to-noise ratio $\delta < 0.25$, and the cited $\Delta m$ is likely over-estimated \citep{Tokovinin2010}.}
\tablenotetext{a}{References: HRCam1 = \cite{Tokovinin2024}; HRCam2 = \cite{Mason2023}; HRCam3 = \cite{Tokovinin2022}; HRCam4 = \cite{Tokovinin2021b}. }
\end{deluxetable*}

Meanwhile, Table \ref{tab:nondetections}, summarizes those high-quality observations for which no resolution of the target into subcomponents has been achieved.  For these targets a 5-$\sigma$ magnitude contrast detection limit has been placed for two different angular separations using the same method as described in \citet[][see their Section 4.3]{Horch2020} for DSSI observations, and in \citet[][see their Section 2.7]{Tokovinin2010} for HRCam observations.

\begin{figure}
    \centering
    \includegraphics[width=\columnwidth]{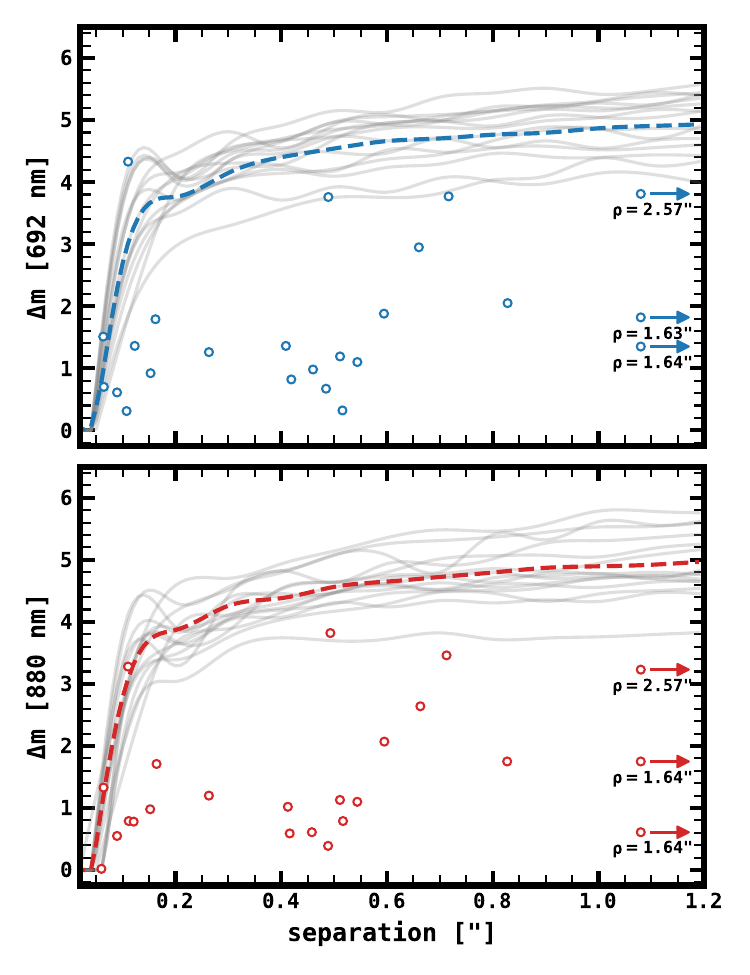}
    \caption{The average detection limits for the targets in our sample from DSSI in the 692 nm channel (top) and in the 880 nm channel (bottom). The gray lines show the detection limits from each of the targets with a high-quality non-detection, and the colored dashed line is the mean of all the detection curves. Targets with speckle-resolved companions are denoted with {\it open circles} (sometimes accompanied by right arrows, the latter to indicate those lying beyond the range of plotted separations).} 
    \label{fig:detectionlimits_widesep}
\end{figure}

Figure \ref{fig:detectionlimits_widesep} shows the detailed detection limit curves as a function of angular separation for all of the HQNDs in Table \ref{tab:nondetections} observed by DSSI and from which the listed 5-$\sigma$ magnitude contrast detection limit values have been derived.  In general these curves follow the same form, but variations arise from observation to observation due to differences in noise and observing conditions (e.g., seeing). The dashed, colored lines represent the mean of these curves.  The points shown represent Q$^+$EBs that were resolved by DSSI observations (Table \ref{tab:speckle measurements}). Companions detected above the mean detection curve simply reflect observations that were of higher quality.

\begin{deluxetable*}{lcDcccccccccc}
\tabletypesize{\scriptsize}
\tablewidth{0pt}
\tablecolumns{9}
\tablecaption{High-Quality Non-Detections (HQNDs)\label{tab:nondetections}}
\tablehead{
\colhead{TIC \#} & \colhead{Gmag} & \multicolumn2c{Date} & & \multicolumn2c{$5\sigma$ Det. Lim., 692nm} & & \multicolumn2c{$5\sigma$ Det. Lim., 880nm} & & \multicolumn3c{5$\sigma$Det. Lim., 824nm}\\
\cline{6-7} \cline{9-10} \cline{12-14}
 &  &  \multicolumn2c{(2000+)} & & \colhead{0.2"} & \colhead{1.0"} & & \colhead{0.2"} & \colhead{1.0"} & & \colhead{$\rho_{min}$} & \colhead{0.15"} & \colhead{1.0"}
} 
\decimals
\startdata         
 52856877 & 10.85 & 22.74073 & & 4.45     & 5.30     & & 4.59     & 5.79     & & $\cdots$ & $\cdots$ & $\cdots$ \\ 
 63459761 & 12.17 & 22.74314 & & 3.62     & 4.86     & & 4.17     & 5.31     & & $\cdots$ & $\cdots$ & $\cdots$ \\ 
 73296637 & 10.75 & 22.87243 & & 3.71     & 4.85     & & 3.66     & 4.56     & & $\cdots$ & $\cdots$ & $\cdots$ \\ 
123098844 & 11.11 & 22.74306 & & 3.64     & 4.55     & & $\cdots$ & $\cdots$ & & $\cdots$ & $\cdots$ & $\cdots$ \\
          &       & 23.35186 & & 4.06     & 5.04     & & 3.67     & 4.78     & & $\cdots$ & $\cdots$ & $\cdots$ \\  
139650665 & 11.23 & 23.79838 & & 4.10     & 4.52     & & 3.48     & 4.33     & & $\cdots$ & $\cdots$ & $\cdots$ \\ 
139944266 & 10.42 & 23.3246  & & $\cdots$ & $\cdots$ & & $\cdots$ & $\cdots$ & & 53.4     & 2.58     & 3.79     \\ 
204698586 & 11.21 & 22.35398 & & $\cdots$ & $\cdots$ & & 3.69     & 4.94     & & $\cdots$ & $\cdots$ & $\cdots$ \\
          &       & 22.36753 & & 3.48     & 4.14     & & 3.63     & 4.45     & & $\cdots$ & $\cdots$ & $\cdots$ \\  
250119205 & 10.42 & 23.5713  & & $\cdots$ & $\cdots$ & & $\cdots$ & $\cdots$ & & 45.8     & 1.03     & 3.76     \\
264402353 & 12.08 & 22.74335 & & 2.97     & 4.39     & & 3.39     & 4.71     & & $\cdots$ & $\cdots$ & $\cdots$ \\ 
271186951 & 11.94 & 23.3246  & & $\cdots$ & $\cdots$ & & $\cdots$ & $\cdots$ & & 65.6     & 1.76     & 2.46     \\
274791367 & 12.51 & 23.4124  & & $\cdots$ & $\cdots$ & & $\cdots$ & $\cdots$ & & 51.4     & 2.41     & 3.21     \\
278352276 & 10.32 & 22.74310 & & 4.15     & 5.27     & & 4.31     & 5.61     & & $\cdots$ & $\cdots$ & $\cdots$ \\ 
284814380 & 12.01 & 22.74071 & & 4.04     & 5.04     & & 3.29     & 5.53     & & $\cdots$ & $\cdots$ & $\cdots$ \\ 
285655079 & 12.86 & 23.5713  & & $\cdots$ & $\cdots$ & & $\cdots$ & $\cdots$ & & 64.6     & 0.64     & 1.83     \\
321471064 & 12.29 & 23.4889  & & $\cdots$ & $\cdots$ & & $\cdots$ & $\cdots$ & & 49.4     & 1.60     & 2.21     \\
370440624 & 11.71 & 23.4124  & & $\cdots$ & $\cdots$ & & $\cdots$ & $\cdots$ & & 41.5     & 2.65     & 4.09     \\
387096013 & 12.14 & 23.4124  & & $\cdots$ & $\cdots$ & & $\cdots$ & $\cdots$ & & 43.4     & 1.77     & 3.20     \\
          &       & 23.9118  & & $\cdots$ & $\cdots$ & & $\cdots$ & $\cdots$ & & 63.0     & 2.23     & 3.21     \\
391620600 & 12.48 & 22.74367 & & 4.10     & 5.42     & & 3.83     & 4.78     & & $\cdots$ & $\cdots$ & $\cdots$ \\ 
434452777 & 10.80 & 22.87270 & & 4.00     & 5.20     & & 3.91     & 4.73     & & $\cdots$ & $\cdots$ & $\cdots$ \\ 
439511833 & 10.59 & 23.4124  & & $\cdots$ & $\cdots$ & & $\cdots$ & $\cdots$ & & 41.5     & 2.23     & 4.55     \\
441794509 & 12.55 & 23.66886 & & 3.69     & 4.40     & & 3.04     & 3.76     & & $\cdots$ & $\cdots$ & $\cdots$ \\ 
454140642 & 10.22 & 22.73819 & & 3.91     & 5.31     & & 3.88     & 5.05     & & $\cdots$ & $\cdots$ & $\cdots$ \\ 
459959916 & 13.22 & 22.74655 & & 4.11     & 5.15     & & 3.89     & 4.70     & & $\cdots$ & $\cdots$ & $\cdots$ \\ 
\enddata
\tablecomments{The 5$\sigma$ detections limits at 824nm are from HRcam, and have been previously published in \cite{Tokovinin2024}. The parameter $\rho_{min}$ is a value reported with HRcam observations and gives the minimum resolvable separation in mas when pairs with $\Delta m$ $<$ 1 mag are detectable.}
\end{deluxetable*}

\subsection{Assessing Astrometric Repeatability}
\label{sec:astrometry_repeat}

As noted in Section \ref{sec:observations}, our previous analysis of the DSSI data collected on the ARC 3.5-m \citep{Davidson2024} have shown good repeatability. This was tested by comparing results obtained contemporaneously between the 692nm and 880nm filter data, including most of the DSSI observations reported here, resulting in an astrometric precision of $2.06 \pm 0.11$ mas. To estimate the astrometric accuracy of DSSI, \citet{Davidson2024} compare the DSSI measured astrometry for binaries with Grade 1 or Grade 2 orbits in the 6th Catalog of Visual Orbits of Binary Stars \citep{Hartkopf2001a}.\footnote{\url{https://crf.usno.navy.mil/wds-orb6}} In that case, Davidson et al. find a larger uncertainty of $\sim$4mas.  Yet another strategy for assessing the astrometric repeatability specifically for the Q$^+$EB observations is to examine the dispersions in the measured position angle and separation across those observations, combining the data from the 692nm and 880nm filters.  One caveat to such a test, however, is that any differential movement between the sources (i.e., if the pair does not share a common proper motion) will inflate the dispersions, leading to overestimates of the uncertainties.

These dispersions in measured position angle ($\sigma_{\theta}$) and separation ($\sigma_{\rho}$) as well as their uncertainties ($\sigma/\sqrt{N}$) are summarized in Table \ref{tab:Repeat_Visits}, and cover nights spanning all of the observing runs. In some cases, e.g., TIC 266771301, TIC 336882813, TIC 389836747, and TIC 470710327, the
dispersions in the measured separations are a close match to the astrometric precision of 2.06 mas reported by \citet{Davidson2024}. On the other hand, the dispersions for some of the other targets range to as much as almost five times higher than the \citet{Davidson2024} value; however that particular result (for TIC 344541836) is strongly affected by the measurements from the first night of data for this source, which were taken in very poor ($\sim$1.5 arcsec) seeing and led to the measured separations from this night's data for this very tight (0.14 arcsec) binary much different from those on the other nights.  The data from the other two nights yield a dispersion in the separation of only 0.8 mas, and a dispersion in the position angle of only 0.25$^{\circ}$.  Meanwhile, the targets with the next two highest dispersions in measured separation are the two with the largest reported separations in our entire Q$^+$EB sample --- TIC 317863971 ($\rho \sim 2.7$ arcsec) and TIC 286470992 ($\rho \sim 1.6$ arcsec).  This may reflect the fact that for larger separations the uncertainty in the pixel scale from one observing run to another will begin to dominate over the measurement repeatability. Note that  \citet[][their Fig.~8]{Davidson2024}  measured the {\it external accuracy} of the DSSI separation measurements to be $\sim$4 mas.  In addition, the Davidson et al. sample, though it included most of the stars reported here, also contained many brighter stars from other programs, and so, in the mean, was a much brighter dataset than the one presented in this work. This difference in mean brightness of the samples alone would lead one to expect a larger dispersion in repeatability of the Q$^+$EB sample.\footnote{As an example for comparison, \citet[][see their Fig.~2a]{Colton2021} obtain a larger dispersion --- 3.1 mas --- in measured separations for fainter binaries (more comparable to the Q$^+$EBs here) using speckle imaging, including with DSSI on the WIYN 3.5-m telescope.}

Overall, the repeatability test summarized in Table \ref{tab:Repeat_Visits} suggests that our measurements should be reliable to at least 8 mas, and much better than this over angular spans smaller than an arcsecond.  However, ultimately the position errors are also a function of contrast and signal-to-noise.

\begin{deluxetable*}{lcccrr}
\tabletypesize{\scriptsize}
\tablewidth{0pt} 
\tablecaption{Dispersions from Repeat Measurements of Resolved Q$^+$EBs\label{tab:Repeat_Visits}}
\tablehead{
\colhead{TIC \#} & \colhead{\# Measures} & \colhead{\# Nights} & \colhead{\# Obs.Runs} & \colhead{$\sigma_{\theta}$}& \colhead{$\sigma_{\rho}$}  \\
\colhead{} & \colhead{} & \colhead{} & \colhead{} & \colhead{($^\circ$)} & \colhead{(")}  
}
\startdata
266771301 &  4  & 2  & 2 & 0.50$\pm$0.18 & 0.0023$\pm$0.0008 \\
286470992 &  4  & 2  & 2 & 0.05$\pm$0.02 & 0.0079$\pm$0.0028 \\
317863971 &  4  & 2  & 2 & 0.10$\pm$0.03 & 0.0078$\pm$0.0028 \\
336882813 &  4  & 2  & 2 & 0.19$\pm$0.07 & 0.0027$\pm$0.0010 \\
344541836 &  6  & 3  & 3 & 0.54$\pm$0.16 & 0.0097$\pm$0.0028 \\
348651800 &  4  & 2  & 2 & 0.21$\pm$0.07 & 0.0066$\pm$0.0023 \\
367448265 &  6  & 3  & 3 & 0.08$\pm$0.02 & 0.0063$\pm$0.0018 \\
389836747 &  4  & 2  & 2 & 0.51$\pm$0.18 & 0.0014$\pm$0.0005 \\
470710327 &  16 & 3  & 2 & 0.13$\pm$0.02 & 0.0018$\pm$0.0003 \\
\enddata
\end{deluxetable*}

\subsection{Assessing Photometric Repeatability}
\label{sec:photometry_repeat}

Repeat observations of sources in Table \ref{tab:speckle measurements} might also be used to gauge photometric reliability, but here more care is required in evaluating repeat observations of, e.g., the $\Delta m$ values in Table \ref{tab:speckle measurements}. 

First, as has been previously shown (see, e.g., Figure 5b of \citealt{Horch2004}), the error in the relative photometry of close binary stars obtained from fringe fitting of speckle imaging data is a strong function of $\Delta m$ itself, with an upturn expected at both small and large $\Delta m$.  An upturn at large $\Delta m$ is intuitive because there is naturally a loss in contrast in the secondary speckle pattern as the difference in the brightness of the two stars increases. However the uncertainty in $\Delta m$ also increases as the two stars become close in brightness because of subtleties in how the magnitude difference is being evaluated via the depth of the interference fringes in the Fourier plane, and because the ultimate photometric uncertainty is very sensitive to small uncertainties in the power spectrum minimum, a problem exacerbated at both small and large $\Delta m$ (see \citealt{Horch2001} for more details).

So, for example, TIC 97356407, which was observed independently by DSSI and HRCam, is a case of a system with a large magnitude difference, apparently some 3 to 4.5 magnitudes depending on wavelength, that would be expected to lie in a regime with $\sigma_{\Delta m} \sim 0.15$ for an average quality observation.  However, this system is also at a declination of $\delta \sim -30^{\circ}$, which required large airmass, large dispersion observations from APO.  Finally, the HRCam observation was unfortunately also of lower quality.  Thus, it is not surprising that there is some disagreement between the reported photometric data for this Q$^+$EB in Table \ref{tab:speckle measurements}.

Second, the speckle photometric measurements are more sensitive to speckle decorrelation, wherein the primary and secondary speckle patterns are not identical.  It has been previously shown (also in \citealt{Horch2004}) that the criterion (seeing)$\times$(separation) $<$ 0.6 arcsec$^2$ serves as a good indicator for isoplanicity, whereas observations for which this criterion is exceeded should be viewed with caution.  Certainly of concern are systems with large inherent separation.

An example is TIC 286470992, which we resolve into two subsystems with a separation of 1.6 arcsec.  As we show below (Sec.~\ref{sec:Gaia_resolve}, Table \ref{tab:Gaia_detections}), {\it Gaia} also resolves this system with a similar separation and a $\Delta G = 1.67$ mag. However, our observations were obtained with seeing exceeding 1.2 arcsec, yielding (seeing)$\times$(separation) $\gtrsim$ 2, and therefore will have suffered from speckle decorrelation. This will skew the derived $\Delta m$ values higher than otherwise, which is consistent with them exceeding the {\it Gaia} $\Delta G$.  According to Figure 9 of \citet{Davidson2024}, for (seeing)$\times$(separation) in the range of 1.96-2.29 (the range of values for our observations of this source), the DSSI $\Delta m$ measurements will be exaggerated over those of {\it Gaia} by some 0.2-1.0 magnitudes, which is consistent with what we see.

To mark those photometric measurements that may suffer inflated $\Delta m$ due to speckle decorrelation, Table \ref{tab:speckle measurements} flags those $\Delta m$ measures that are derived from observations with (seeing)$\times$(separation) $>$ 0.6 arcsec$^2$. One should be aware that some systems --- e.g., TIC 317863971 --- will almost always be in this special regime no matter the quality of the seeing by virtue of the wide separation of the subcomponents (in this particular case, 2.6 arcsec). 

A third situation of concern are observations of low $S/N$, where the photometric measurements are simply of lower quality.  An example is TIC 336882813, a fainter source ($V \sim 12$) for which one set of observations was taken in cloudy conditions.  Table \ref{tab:speckle measurements} shows a larger spread in $\Delta m$, but a spread not inconsistent with expectations for the quality of data, $\sigma_{\Delta m} \sim 0.2$ mag.   Similarly, the first visit to TIC 344541836 was taken in particularly bad seeing (as noted in Sec.~\ref{sec:astrometry_repeat}),  so there is no surprise that the photometric data for this evening is greatly disparate from those of the other nights.

To indicate observations of lower quality, we have marked DSSI $\Delta m$ values in Table \ref{tab:speckle measurements} with an asterik (*) and HRCam $\Delta m$ values with a colon (:). For DSSI, as previously mentioned, these are defined as observations having (seeing)$\times$(separation) $> 0.6$ arcsec$^{2}$, and for HRcam these are defined as observations having a S/N $\delta < 0.25$ \citep{Tokovinin2010}

\subsection{Comparison of Speckle Imaging Resolution of Q$^+$EBs to those by Gaia DR3}
\label{sec:Gaia_resolve}

The resolution of {\it Gaia} (EDR3), if defined by the minimum angular separation between distinct catalog sources (i.e., objects with different {\tt source\_id}), is 0.18 arcsec; however, in practice a small fraction of sources with separations below 0.6 arcsec are resolved and these generally only have 2-parameter solutions \citep[][Section 5.2]{Lindegren2021}.  In reality, {\it Gaia}'s resolving power depends on a number of factors, including the orientation of the position angle of a binary star relative to the dominant scan direction of the satellite on that source as well as the relative brightness of the stars.  The latter effect alone creates a large variation in resolution, from $\sim$0.5 arcsec for stars of equal magnitude to as poor as $\sim1$ arcsec to $\sim5$ arcsec for stars differing by $\Delta G = 5$ to $10$ mag \citep[][Fig.~30]{GaiaCollab2021}.  \citet[][Section 2.3]{Fabricius2021} tackle the question of {\it Gaia} EDR3 completeness on small scales more directly.  Overall they find {\it Gaia} EDR3 to show a great improvement over {\it Gaia} DR2 in terms of small-scale completeness.  In an investigation of a dense field in the Galactic plane EDR3 was shown to be highly complete above 1.5 arcsec source separations.  However, below 1.5 arcsec completeness begins to drop off to about 80\% at 0.7 arcsec, at which point completeness drops off precipitously, with almost no close stellar pairs resolved below 0.4 arcsec (except expected spurious detections). A second check of {\it Gaia} EDR3 against the Washington Double Star Catalog \citep{Mason2001} shows similar, albeit slightly better, results, but still a strong decline in completeness on subarcsecond scales, with 50\% incompleteness at 0.5 arcsec.  

Our sample of resolved, but compact Q$^+$EBs offers yet another opportunity to investigate {\it Gaia} performance on small scales, but also to evaluate its effectiveness in exploring the growing catalog of Q$^+$EB sources. While {\it Gaia} will not be able to compete with diffraction-limited imaging by large aperture telescopes in the realm of resolving compact Q$^+$EBs, it may still prove useful at identifying the more widely separated systems in advance of such large telescope observing.  Even more important, since we are so dependent on {\it Gaia} trigonometric parallaxes for distance estimation, is to understand how the presence of subarcsecond image structure due to Q$^+$EB multiplicity may effect {\it Gaia} astrometry. 

To undertake our analysis, we first cross-matched the K22 sample with {\it Gaia} DR3 by querying the latter for all sources within 5 arcsec of the source identified (by SIMBAD) with each TIC ID in the former. We find that {\it Gaia} DR3 contains a total of 35 K22 sources where there is one or more companion within 5.0 arcsec. Position angle ($\theta$), angular separation ($\rho$), $\Delta G$, and individual magnitudes for sources within 3.0 arcsec of the {\it Gaia} coordinates for each TIC source, are presented in Table \ref{tab:Gaia_detections}. Nine of these fifteen sources fall in our Table \ref{tab:speckle measurements}, and Table \ref{tab:Gaia_detections} includes our measurements for these sources for comparison.  For seven of those nine, there is reasonable to good agreement between the {\it Gaia} and speckle measurements for the position angle and angular separation of the resolved sources. There is more variation in the magnitude differences, but this is to be expected, given the variation in the passbands.

\startlongtable
\begin{deluxetable*}{lccccccl}
\tabletypesize{\scriptsize}
\tablewidth{0pt} 
\tablecaption{Measurements of {\it Gaia} Resolved Q$^+$EBs\label{tab:Gaia_detections}}
\tablehead{
\colhead{TIC \#} & \colhead{$\theta$}& \colhead{$\rho$} & \colhead{Primary $G$} & \colhead{Secondary $G$} & \colhead{$\Delta$$m$} & \colhead{Source} \\
\colhead{} & \colhead{($^\circ$)} & \colhead{(")} & \colhead{(mag)} & \colhead{(mag)} & \colhead{(mag)} & 
} 
\startdata 
TIC75740921  &  51.4 & 2.4462 & 13.22 & 17.77 & 4.55 & {\it Gaia}\\
             &  95.9 & 0.3986 &       &       & 0.4  & HRCam - 824nm\\ 
\hline
TIC123098844\tablenotemark{a} & 334.3 & 1.6128 & 11.11 & 16.90 & 5.79 & {\it Gaia}\\
\hline
TIC130276377\tablenotemark{b} & 123.9 & 2.8964 & 12.00 & 18.08 & 6.08 & {\it Gaia}\\
\hline
TIC168789840 & 263.7 & 0.3745 & 11.67 & 12.01 & 0.34 & {\it Gaia}\\
             & 257.7 & 0.4230 &       &       & 0.3  & HRCam - 824nm\\
             & 257.6 & 0.4233 &       &       & 0.3  & HRCam - 824nm\\
             & 257.6 & 0.4235 &       &       & 0.3  & HRCam - 824nm\\
             & 257.7 & 0.4243 &       &       & 0.3  & HRCam - 517nm\\
             & 258.1 & 0.4221 &       &       & 0.3  & HRCam - 824nm\\
             & 258.1 & 0.4229 &       &       & 0.2  & HRCam - 517nm\\
             & 257.8 & 0.4228 &       &       & 0.4  & HRCam - 824nm\\
             & 258.9 & 0.4190 &       &       & 0.82 & DSSI - 692nm\\
             & 257.9 & 0.4160 &       &       & 0.59 & DSSI - 880nm\\
             & 258.1 & 0.4231 &       &       & 0.2  & HRCam - 824nm\\
             & 258.4 & 0.4234 &       &       & 0.3  & HRCam - 824nm\\
\hline
TIC200094011 & 256.0 & 0.6789 &  9.71 & 10.58 & 0.87 & {\it Gaia}\\
             & 208.6 & 0.5439 &       &       & 1.10 & DSSI - 692nm\\
             & 208.7 & 0.5438 &       &       & 1.10 & DSSI - 880nm\\ 
             & 208.4 & 0.5435 &       &       & 1.0  & HRCam - 824nm\\
\hline
TIC219469945\tablenotemark{b} & 177.6 & 2.4383 & 12.42 & 16.73 & 4.31 & {\it Gaia}\\
\hline
TIC239872462 &  11.4 & 1.6330 & 11.29 & 12.51 & 1.22 & {\it Gaia}\\
             &  11.1 & 1.6365 &       &       & 1.35 & DSSI - 692nm\\
             &  11.2 & 1.6385 &       &       & 0.61 & DSSI - 880nm\\
\hline
TIC266771301 & 215.5 & 0.5046 & 12.26 & 12.88 & 0.62 & {\it Gaia}\\
             & 215.2 & 0.5158 &       &       & 0.32 & DSSI - 692nm\\
             & 215.0 & 0.5168 &       &       & 0.79 & DSSI - 880nm\\
             & 214.9 & 0.5177 &       &       & 0.00 & DSSI - 692nm\\
             & 216.0 & 0.5124 &       &       & 0.00 & DSSI - 880nm\\
\hline
TIC269811101\tablenotemark{b} & 129.1 & 2.7642 & 14.15 & 20.00 & 5.85 & {\it Gaia}\\
\hline
TIC285681367\tablenotemark{b} & 179.5 & 2.4540 & 12.13 & 19.73 & 7.60 & {\it Gaia}\\
\hline
TIC286470992 & 198.2 & 1.6340 & 10.11 & 11.78 & 1.67 & {\it Gaia}\\
             & 198.1 & 1.6336 &       &       & 1.82 & DSSI - 692nm\\
             & 198.1 & 1.6353 &       &       & 1.75 & DSSI - 880nm\\
             & 198.1 & 1.6501 &       &       & 2.26 & DSSI - 692nm\\
             & 198.0 & 1.6453 &       &       & 2.00 & DSSI - 880nm\\
\hline
TIC307119043 & 157.6 & 0.8337 & 10.01 & $\cdots$ & $\cdots$ & {\it Gaia}\\
             & 157.7 & 0.8281 &       &       & 2.05 & DSSI - 692nm\\
             & 157.7 & 0.8273 &       &       & 1.75 & DSSI - 880nm\\
\hline
TIC317863971 & 184.7 & 2.5449 & 10.32 & 13.93 & 3.61 & {\it Gaia}\\
             & 184.6 & 2.5691 &       &       & 3.81 & DSSI - 692nm\\
             & 184.7 & 2.5669 &       &       & 3.23 & DSSI - 880nm\\
             & 184.5 & 2.5793 &       &       & 3.92 & DSSI - 692nm\\
             & 184.5 & 2.5830 &       &       & 3.36 & DSSI - 880nm\\
\hline
TIC375325607 &  80.9 & 0.4787 & 13.14 & 13.25 & 0.11 & {\it Gaia}\\
             &  79.3 & 0.4847 &       &       & 0.67 & DSSI - 692nm\\
             &  79.5 & 0.4886 &       &       & 0.39 & DSSI - 880nm\\
\hline
TIC434452777\tablenotemark{a} & 288.7 & 2.6826 & 10.80 & 18.76 & 7.96 & {\it Gaia}\\
\enddata
\tablecomments{}
\tablenotetext{a}{HQND result with DSSI.}
\tablenotetext{b}{Observations with DSSI had lower SNR, and are not included in the present study.}
\end{deluxetable*}

For TIC75740921, it is clear that {\it Gaia} is resolving a different target than was resolved by HRCam.  In the first place, the angular separations of the reported pairs are very different, and it is unlikely that {\it Gaia} would have resolved the 0.40 arcsec sources resolved by HRCam, while the $G$ = 17.77 source reported by {\it Gaia} DR3 is beyond the magnitude range of typical HRCam observations.  On the other hand, for the case of TIC200094011 the reported separations and position angles between the sources measured by HRCam and DSSI are in good agreement, but both differ somewhat from {\it Gaia}'s reported values. It's possible that the differences may be due to the fact that this object, reported to be a young stellar object, is apparently embedded in a reflection nebulae, Messier 78 (NGC 2068), which creates a complicated background.

There are six sources in Table \ref{tab:Gaia_detections} with no corresponding speckle observations yielding a resolution into two sources. Four of these sources --- TIC130276377, TIC219469945, TIC269811101, and TIC285681367 --- have been observed with DSSI at APO, however the observations to date have had low SNR and therefore no results are reported in Tables \ref{tab:speckle measurements} and \ref{tab:nondetections}. The remaining two sources, TIC123098844 and TIC434452777, have been observed with DSSI yet appear in Table \ref{tab:nondetections} as unresolved.  This is not surprising, because our speckle observations on 4-m class telescopes would not be expected to detect the $G = 16.90$ and $G = 18.76$ secondaries reported by {\it Gaia}.  

\begin{figure}
{\includegraphics[width=\columnwidth]{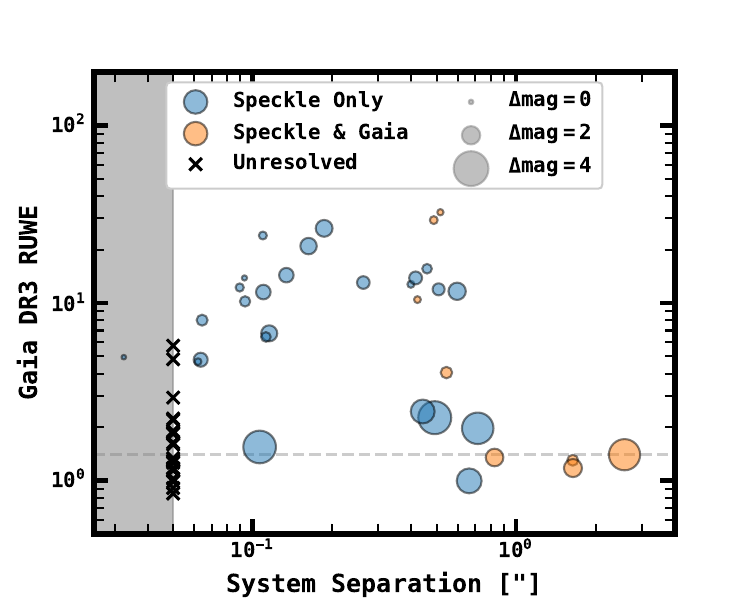}}
         \caption{
The {\it Gaia} DR3 RUWE as a function of the separation of Q$^+$EBs. The {\it  orange circles} represent the eight Q$^+$EBs that are resolved both by speckle imaging and {\it Gaia}, while the 
{\it blue circles} represent those systems resolved only by 
speckle imaging. The stack of {\it crosses} shown at a separation of 0.05 arcsec represents the high quality non-detections in Table \ref{tab:nondetections}, assigned a separation corresponding to the approximate resolution limit for our speckle imaging. For the resolved systems, the size of the symbol represents the relative magnitude difference of the two components. The {\it Gaia}-resolved systems show a clear trend where those with smaller separations have smaller brightness contrasts, but larger RUWE. The {\it shaded region} to the left indicates the area below the approximate diffraction limit of the speckle imaging, which is 50 mas for both telescopes used, accounting for the wavelengths adopted for observations at each site. 
\label{fig:Sep_RUWE}}
\end{figure}

Overall, the results presented in Table \ref{tab:Gaia_detections} suggest that we can generally believe {\it Gaia} resolution results for Q$^+$EB targets, when they exist --- however, they exist only under restricted circumstances. Figure \ref{fig:Sep_RUWE} summarizes the distribution of separations of systems resolved by {\it Gaia} and speckle imaging.  The ordinate of the plot shows the {\it Gaia} DR3 renormalised unit weight error (RUWE) value for the primary (i.e., brightest) source associated with each TESS source, while the abscissa of the plot shows the average speckle-determined separation. The size of the symbol represents the average speckle magnitude difference (from values in Table 1) of the resolved components of the Q$^+$EB. Sources unresolved via speckle imaging (i.e., the HQNDs of Table \ref{tab:nondetections}) are included in Figure \ref{fig:Sep_RUWE} as the black crosses placed at our approximate speckle imaging diffraction limit, or $\sim0.05$ arcsec. A first trend made immediately evident by Figure \ref{fig:Sep_RUWE} is that close to its diffraction limit {\it Gaia} can only resolve close sources that have a small magnitude difference ($\lesssim$1 mag at $\sim$0.5 arcsec), and is unable to resolve sub-arcsecond binaries with magnitude differences exceeding 1 magnitude. This is in line with the trend mentioned above --- i.e., that {\it Gaia}'s close source resolution is particularly dependent on the relative brightness of the sources \citep{GaiaCollab2021}.

Perhaps more concerning is the trend that the {\it Gaia} RUWE for almost all of the sources resolved by only speckle imaging is high.  This shows that the {\it Gaia} parallaxes for those sources that {\it Gaia} does not resolve but that speckle imaging does resolve are not going to be very reliable.  An RUWE $\ge 1.4$ is often adopted as a signature for the presence of an underlying binary, even if unresolved by {\it Gaia}, but \citet{Stassun2021}  show that even sources with RUWE $<$ 1.4 can still be binaries. However, it is evident that for those Q$^+$EBs that speckle imaging resolves into two sources with a large magnitude difference the RUWE was systematically lower, which suggests that the {\it Gaia} solution may have been less influenced by the presence of the fainter source. Overall, the conclusion one draws from this analysis is that {\it Gaia} trigonometric parallaxes may not provide a wholly reliable tool for gauging distances to the TESS-identified Q$^+$EBs.

\subsection{Estimating Q$^+$EB System Distances and Sizes}
\label{sec:distances}

\begin{figure}[h]
{\includegraphics[width=\columnwidth]{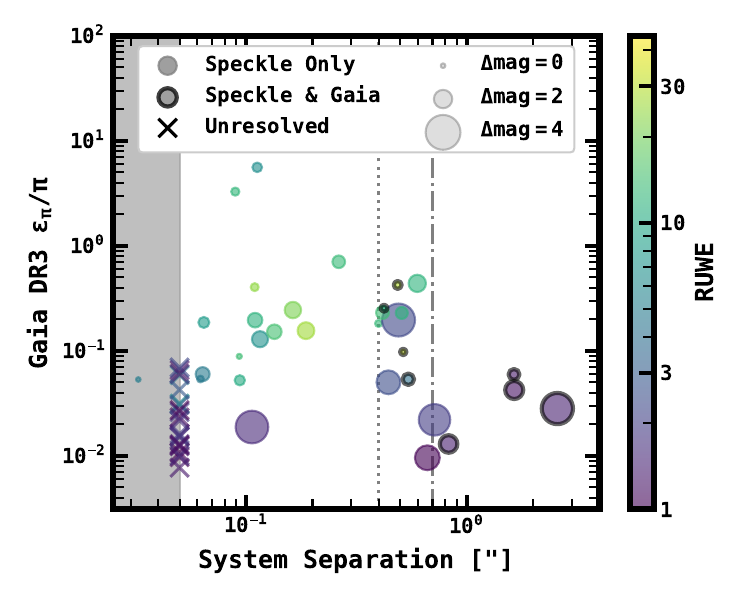}}
         \caption{Relative parallax error,   $\epsilon_{\pi}/{\pi}$, as a function of Q$^+$EB system angular separation. Similar to Figure \ref{fig:Sep_RUWE}, the vertical stack of {\it crosses} at separation of 0.05 arcsec represents the HQNDs, assigned a separation corresponding to the approximate resolution limit for our speckle imaging; the {\it shaded region} to the left indicates the area below this formal diffraction limit. Q$^+$EBs resolved by both our speckle imaging and {\it Gaia} are denoted with a {\it black open circle} (albeit sometimes shaded with interior coloring). As in Fig.~\ref{fig:Sep_RUWE}, the size of the points corresponds to the relative magnitude difference of the two components as determined from our speckle imaging. Points with a purplish hue correspond to objects with {\it Gaia} RUWE $\lesssim$ 1.4, which is where {\it Gaia} is able to well-fit a single-star astrometric solution to describe its observations, and is often adopted as a delimiter above which sources are often suspected of multiplicity \citep{Fabricius2021}. The {\it dotted} and {\it dot-dashed lines} at separations of 0.4 and 0.7 arcsec correspond to the approximate angular resolution achieved by {\it Gaia} DR2 and the visual binary completeness limit of \textbf{{\it Gaia}} DR3, respectively \added{\citep{GaiaCollab2018,Fabricius2021}}
         \label{fig:Gaia_errors}}
\end{figure}

The latter statement is more or less borne out by the actual {\it Gaia} parallax data.  Figure \ref{fig:Gaia_errors} shows {\it Gaia} parallax error, normalized to parallax, versus the angular separations of the Q$^+$EB sources (or the upper limits of the HQNDs, summarily plotted at the 0.05 arcsec ground-based diffraction limit
) derived from our speckle-imaging. As may be seen, for those resolved systems with separations $\gtrsim0.7$ arcsec, the {\it Gaia} DR3 visual binary completeness limit, the fractional {\it Gaia} parallax errors are small.  All members of this well-separated subsample, save TIC 427092089 --- seen straddling the $0.7$ arcsec threshold with a large $\Delta m$ --- are resolved by {\it Gaia}. However, in the transition region of angular separations between the {\it Gaia} DR3 visual binary detection limit and the 0.4 arcsec {\it Gaia} DR2 effective diffraction limit, the fractional parallax errors start increasing rapidly. Sources with similar brightness remain resolved by {\it Gaia} in this range, but those with significant magnitude differences are consistently unresolved by {\it Gaia}.  Finally, below the 0.4 arcsec limit, no targets are resolved by {\it Gaia}, regardless of relative magnitude.

The net result is that for the majority of the Q$^+$EBs that we resolve with speckle imaging, the {\it Gaia} trigonometric parallaxes are unreliable due to the intrinsic subarcsecond  morphological complexity of the sources, which makes it jeopardous to convert the angular separations into linear dimensions for the resolved systems.  
Fortunately, however, for the most compact Q$^+$EBs --- including those that remain unresolved with our speckle imaging --- the {\it Gaia} parallaxes have low fractional errors and the solutions by and large have low RUWE.  We provide these distances and distance errors, based on the \citet{Bailer2021}  {\it Gaia} parallax-based, geometric distances, along with the associated {\it Gaia} DR3 RUWE values, in Table \ref{tab:upperlimit}. 

By adopting the resolution limit for the appropriate telescope and filter combination with which each HQND was observed, we can thereby place an upper limit to the projected linear size of each system as viewed from Earth (on those nights the speckle imaging was obtained), under the assumption of favorable contrast ratios. Such limits are also given in Table \ref{tab:upperlimit}. Of course, the strongest limit will always be given by the bluest (i.e., 692nm) filter.  As may be seen, these blue upper limits range from as high as 175 au for TIC 387096013 to as small as 18 au for TIC 454140642.  In the case of the latter system, this is still a quite generous upper limit, as \citet{Kostov2021} have already extensively explored this ``compact, coplanar, quadruple-lined quadruple star system consisting of two eclipsing binaries'' and shown that it's maximum projected separation as seen from Earth would be $\sim2$ au (see their Fig.~14).  Moreover, 
two of the systems (TIC 52856877 and TIC 278352276) were noted by K22 to be exhibiting prominent eclipse timing variations (ETVs), which is a good indicator that one of the eclipsing binaries within the system is being perturbed by close dynamical interaction with the other stars in the system (see Section \ref{sec:wide_ETVs}). Thus, it is likely that the upper limits in Table \ref{tab:upperlimit} may be quite generous for many of the systems.

Meanwhile, Table \ref{tab:linearsize} (second column) gives the {\it Gaia}-based \citep{Bailer2021} geometric distances for the speckle-resolved systems from Table \ref{tab:speckle measurements}, regardless of the $\epsilon_{\pi}/{\pi}$ fractional error and the associated RUWE values, as well as the implied projected linear separations of the speckle-resolved components (fifth column), based on the unweighted mean-measured angular separation (fourth column).  In those cases where Gaia was able to resolve the same stars {\it and} individual Bailer-Jones distances are available, the second distance is provided in the third column of Table \ref{tab:linearsize} and that second distance is used as another estimator of the linear separation in column six. Given the comments at the end of Section \ref{sec:Gaia_resolve} about the {\it Gaia} parallaxes for these particular sources, the values in Table \ref{tab:linearsize} are provided only as a guide and with a very clear admonition of {\it caveat emptor}.  As far as the numbers can be trusted, most of the resolved systems have linear projected separations that appear to be quite large --- hundreds to thousands of au --- although several systems --- e.g., TIC 97356407 at 32 au, TIC 146810480 at 35 au, TIC 328181241 at 38 au, and TIC 399492452 at 30 au --- have fairly reliable parallaxes that yield system sizes of order that of the solar system.

\begin{deluxetable*}{rrrrrcr}
\def\arraystretch{1.5}
\tabletypesize{\scriptsize}
\tablecaption{Upper Limit of Projected System Linear Size for the High Quality Non-Detections (HQNDs)
\label{tab:upperlimit}}
\tablehead{
\multicolumn{3}{c}{} & \multicolumn{2}{c}{$D$ = 3.5-m} & & \colhead{$D$ = 4.1-m}\\
\cline{4-5} \cline{7-7}
\colhead{TIC \#} & \colhead{RUWE} & \colhead{r (pc)} & \colhead{a$_{\rm 692nm}$ (au)} &  \colhead{a$_{\rm 880nm}$ (au)} &  & \colhead{a$_{\rm 824nm}$ (au)}
}
\startdata 
52856877 & 2.19 & $831_{-22}^{+23}$ & $41.4_{-1.1}^{+1.2}$ & $52.6_{-1.4}^{+1.5}$ & & $\cdots$\\
63459761 & 1.26 & $2017_{-117}^{+133}$ & $100.3_{-5.8}^{+6.6}$ & $127.6_{-7.4}^{+8.4}$ & & $\cdots$\\
73296637 & 1.18 & $585_{-7}^{+6}$ & $29.1_{-0.4}^{+0.3}$ & $37.0_{-0.5}^{+0.4}$ & & $\cdots$\\
123098844 & 1.34 &$756_{-10}^{+9}$ & $37.6_{-0.5}^{+0.5}$ & $47.8_{-0.6}^{+0.6}$ & & $\cdots$\\
139650665 & 5.75 & $257_{-8}^{+7}$ & $12.8_{-0.4}^{+0.4}$ & $16.2_{-0.5}^{+0.5}$ & & $\cdots$\\
139944266 & 1.02 & $1026_{-16}^{+16}$ & $\cdots$ & $\cdots$ & & $51.9_{-0.8}^{+0.8}$\\
204698586 & 4.82 & $551_{-30}^{+27}$ & $27.4_{-1.5}^{+1.4}$ & $34.9_{-1.9}^{+1.7}$ & & $\cdots$\\
250119205 & 1.84 & $665_{-17}^{+17}$ & $\cdots$ & $\cdots$ & & $33.6_{-0.9}^{+0.9}$\\
264402353 & 0.99 & $1003_{-8}^{+8}$ & $49.9_{-0.4}^{+0.4}$ & $63.5_{-0.5}^{+0.5}$ & & $\cdots$\\
274791367 & 2.23 & $2969_{-191}^{+204}$ & $\cdots$ & $\cdots$ & & $150.1_{-9.6}^{+10.3}$\\
278352276 & 0.84 & $744_{-7}^{+7}$ & $\cdots$ & $\cdots$ & & $37.6_{-0.4}^{+0.3}$\\
284814380 & 0.91 & $2422_{-60}^{+56}$ & $120.5_{-3.0}^{+2.8}$ & $153.3_{-3.8}^{+3.5}$ & & $\cdots$\\
321471064 & 1.87 & $816_{-12}^{+13}$ & $\cdots$ & $\cdots$ & & $41.3_{-0.6}^{+0.7}$\\
370440624 & 1.90 & $664_{-10}^{+11}$ & $\cdots$ & $\cdots$ & & $33.6_{-0.5}^{+0.6}$\\
387096013 & 1.28 & $3455_{-165}^{+156}$ & $\cdots$ & $\cdots$ & & $174.8_{-8.4}^{+7.9}$\\
391620600 & 1.59 & $456_{-4}^{+5}$ & $22.7_{-0.2}^{+0.3}$ & $28.8_{-0.3}^{+0.3}$ & & $\cdots$\\
434452777 & 2.93 & $663_{-24}^{+27}$ & $33.0_{-1.2}^{+1.3}$ & $42.0_{-1.5}^{+1.7}$ & & $\cdots$\\
439511833 & 1.14 & $855_{-16}^{+21}$ & $\cdots$ & $\cdots$ & & $43.2_{-0.8}^{+1.1}$\\
441794509 & 1.14 & $882_{-11}^{+12}$ & $43.9_{-0.5}^{+0.6}$ & $55.8_{-0.7}^{+0.7}$ & & $\cdots$\\
454140642 & 1.30 & $356_{-3}^{+3}$ & $17.7_{-0.1}^{+0.2}$ & $22.5_{-0.2}^{+0.2}$ & & $\cdots$\\
459959916 & 1.63 & $1028_{-19}^{+16}$ & $51.2_{-0.9}^{+0.8}$ & $65.1_{-1.2}^{+1.0}$ & & $\cdots$\\
\enddata
\tablecomments{Distances in the third column are based on  {\it Gaia} parallax geometric distances sourced from \citet{Bailer2021}.}
\end{deluxetable*}

\begin{deluxetable*}{rlllrr}
\def\arraystretch{1.5}
\tabletypesize{\scriptsize}
\tablecaption{Estimated Projected Linear Separation of 4-m Class-Telescope-Resolved Q$^+$EBs  \label{tab:linearsize}}
\tablehead{
\colhead{TIC \#} & \colhead{r$_{1}$ (pc)} & \colhead{r$_{2}$ (pc)} & \colhead{$\rho$ (")} & \colhead{a$_{1}$ (au)} & \colhead{a$_{2}$ (au)}\\
} 
\startdata
25818450  & $820_{-53}^{+53}$      & $\cdots$              & 0.0636 & $52.2_{-3.4}^{+3.4}$         & $\cdots$\\
45160946  & $481_{-67}^{+110}$     & $\cdots$              & 0.1869 & $89.8_{-12.6}^{+20.6}$       & $\cdots$\\
75740921  & $865_{-187}^{+259}$    & $\cdots$              & 0.3986 & $344.7_{-74.5}^{+103.3}$     & $\cdots$\\
78568780  & $3236_{-1279}^{+3399}$ & $\cdots$              & 0.2632 & $851.6_{-336.6}^{+894.5}$    & $\cdots$\\
89278612  & $5254_{-1765}^{+2779}$ & $\cdots$              & 0.0894 & $469.7_{-157.8}^{+248.4}$    & $\cdots$\\
97356407  & $300_{-5}^{+4}$        & $\cdots$              & 0.1064 & $31.9_{-0.6}^{+0.5}$         & $\cdots$\\
146810480 & $376_{18}^{+22}$       & $\cdots$              & 0.0937 & $35.2_{-1.7}^{+2.0}$         & $\cdots$\\
161043618 & $1815_{-388}^{+593}$   & $\cdots$              & 0.1095 & $198.8_{-42.5}^{+64.9}$      & $\cdots$\\
168789840 & $623_{-125}^{+156}$    & $*$                   & 0.4175 & $260.2_{-52.3}^{+65.0}$      & $*$     \\
200094011 & $488_{-24}^{+27}$      & $*$                   & 0.5437 & $265.2_{-12.9}^{+14.7}$      & $*$     \\
201310151 & $950_{-98}^{+113}$     & $\cdots$              & 0.1157 & $109.9_{-11.3}^{+13.0}$      & $\cdots$\\
239872462 & $1765_{-103}^{+119}$   & $1658_{-84}^{+68}$    & 1.6375 & $2889.4_{-168.4}^{+195.6}$   & $2714.4_{-136.9}^{+109.2}$\\
255532033 & $4696_{-1858}^{+2417}$ & $\cdots$              & 0.1124 & $527.9_{-208.8}^{+271.7}$    & $\cdots$\\
257776944 & $1113_{-198}^{+259}$   & $\cdots$              & 0.1099 & $122.4_{-21.8}^{+28.5}$      & $\cdots$\\
266771301 & $216_{-20}^{+24}$      & $3034_{-934}^{+1310}$ & 0.5157 & $111.6_{-10.2}^{+12.5}$      & $1564.6_{-481.6}^{+675.3}$\\
283940788 & $4185_{-475}^{+852}$   & $\cdots$              & 0.4910 & $2055.0_{-233.2}^{+418.5}$   & $\cdots$\\
286470992 & $2120_{-83}^{+96}$     & $1905_{-104}^{+123}$  & 1.6411 & $3479.4_{-137.0}^{+158.3}$   & $3126.4_{-170.8}^{+201.4}$\\
300446218 & $467_{-40}^{+42}$      & $\cdots$              & 0.0932 & $43.6_{-3.8}^{+3.9}$         & $\cdots$\\
307119043 & $465_{-6}^{+7}$        & $486_{-20}^{+20}$     & 0.8277 & $384.9_{-5.4}^{+5.6}$        & $402.0_{-17.0}^{+16.5}$ \\
317863971 & $716_{-21}^{+22}$      & $708_{-9}^{+9}$       & 2.5746 & $1842.5_{-53.5}^{+56.4}$     & $1822.7_{-23.6}^{+23.9}$\\
322727163 & $1098_{-325}^{+435}$   & $\cdots$              & 0.1631 & $179.2_{-53.0}^{+70.9}$      & $\cdots$\\
328181241 & $607_{-33}^{+32}$      & $\cdots$              & 0.0622 & $37.8_{-2.0}^{+2.0}$         & $\cdots$\\
336882813 & $2867_{-948}^{+1220}$  & $\cdots$              & 0.5970 & $1711.7_{-565.8}^{+728.1}$   & $\cdots$\\
344541836 & $483_{-70}^{+95}$      & $\cdots$              & 0.1344 & $65_{-9.4}^{+12.7}$          & $\cdots$\\
348651800 & $814_{-169}^{+274}$    & $\cdots$              & 0.4154 & $338.3_{-70.2}^{+113.9}$     & $\cdots$\\
357810643 & $562_{-34}^{+28}$      & $\cdots$              & 0.4419 & $248.2_{-15.0}^{+12.4}$      & $\cdots$\\
367448265 & $327_{-3}^{+2}$        & $\cdots$              & 0.6631 & $217.1_{-2.0}^{+1.6}$        & $\cdots$\\
375325607 & $5741_{-2552}^{+2119}$ & $*$                   & 0.4867 & $2794.0_{-1241.8}^{+1031.2}$ & $*$     \\
389836747 & $*$                    & $\cdots$              & 0.1536 & $*$                          & $\cdots$\\
399492452 & $930_{-49}^{+52}$      & $\cdots$              & 0.0325 & $30.2_{-1.6}^{+1.7}$         & $\cdots$\\
427092089 & $1146_{-22}^{+23}$     & $\cdots$              & 0.7146 & $818.8_{-15.6}^{+16.4}$      & $\cdots$\\
438226195 & $6846_{-2057}^{+2439}$ & $\cdots$              & 0.4590 & $3142.3_{-944.3}^{+1119.4}$  & $\cdots$\\
461614217 & $1847_{-323}^{+495}$   & $\cdots$              & 0.0644 & $118.9_{-20.8}^{+31.8}$      & $\cdots$\\
470710327 & $1263_{-393}^{+1348}$  & $\cdots$              & 0.5078 & $641.6_{-199.7}^{+684.6}$    & $\cdots$\\
\enddata
\tablecomments{Sources that were resolved by Gaia but that have no available DR3 parallax are denoted with an asterisk (*).}
\end{deluxetable*}

\subsection{Systems of Special Interest}
\label{sec:special}

\subsubsection{The Sextuple System TIC 168789840}

\citet{Powell2021} first reported the discovery of the sextuply eclipsing system TIC 168789840.  As part of their analysis, speckle imaging of the system was obtained with HRCam on the SOAR telescope over 3 nights (2020.8233 -- 2020.9238) and in the $V$ and $I$ filters.  The AC subsystem binary components were clearly separated from the B subsystem binary with a mean separation of 0.42 arcsec and mean position angle of 257.7 deg.  These values correspond very well with our 2022.73817 DSSI-measured values of 0.42 arcsec and 258.4 deg. However, the measured magnitude differences in Powell et al. are  $\Delta V=0.31$ mag and $\Delta I =0.28$ mag, about half the size of the magnitude difference in either the 692nm or 880nm filters reported in Table \ref{tab:speckle measurements} --- though the effective wavelength of the Cousins $I$ filter falls roughly in between either of these DSSI passbands.  

We investigated the DSSI results for this source in greater detail. While those observations were taken in good seeing (measured to be $\sim0.85$ arcsec in the two passbands), the $S/N$ of the DSSI images were low.  This target has a declination of $-32^{\circ}$, and was observed at an airmass of $\sim$2.4, which would have significantly increased the dispersion.   Because TIC 168789840 is relatively faint, three 1000-frame sequences were obtained with DSSI, and it was found that, when reduced independently, one of the sets gives a much larger $\Delta m$ at 692 nm than the other two sets, and this outlier sequence has undoubtedly influenced our total result in Table \ref{tab:speckle measurements} to be high for this filter.  It is not clear whether the $\Delta m$ result for 880 nm is also exaggerated, but the larger aperture of the telescope and the greater consistency of the multiple observations taken on SOAR suggest that those speckle photometric results may be more trustworthy.

\subsubsection{TIC 266771301}
\label{sec:TIC 266771301}
Table \ref{tab:linearsize} contains five Q$^+$EBs where both speckle-resolved components have parallaxes provided by Gaia.  In four of the cases, the \citet{Bailer2021} distances derived from these parallaxes are generally in good agreement --- only in the case of TIC 286470992 are the distances similar, but the 1-$\sigma$ errors do not overlap.  However, in the case of the system TIC 266771301 the parallaxes provided by Gaia for the two resolved subsystems do not at all agree.  The resulting disagreement in the derived distance is vast, suggesting that the resolved components are not related.  In this case it is possible that (a) the farther source at $\sim 3$ kpc is the Q$^+$EB, placing an upper limit of $\sim 150$ au on the linear extent of the unresolved component; (b) TIC 266771301 consists of two unrelated EBs that just happen to be along the same line of site; or (c) the closer source is the Q$^+$EB, in which case the upper limit on the angular size of the system would be 10.8 au, making it the {\it most compact} example in our entire sample. The example of TIC 266771301 points out the value of follow-up observations, such as speckle imaging during eclipses (\S \ref{sec:SIDEs}), to disambiguate system architectures.

\subsubsection{Systems With Eclipse Timing Variations}
\label{sec:wide_ETVs}

Among the Q$^+$EBs observed in the present study are several that K22 noted to have ETVs. The presence of ETVs indicates influence on an EB of a perturbing force due to a quite proximate star or stars.  It is therefore interesting to review these several ETV cases in the context of the speckle imaging results to check for consistencies.

{\bf TIC 52856877:} This system was noted by K22 to have ``prominent secondary ETVs''.  As shown in Tables \ref{tab:nondetections} and \ref{tab:upperlimit}, our DSSI observations of this system yield an HQND result with an upper limit on the size of the Q$^+$EB based on the 692nm filter of about 41au.  Our HQND result for this target is consistent with the notion that the secondary ETVs are being induced by the primary EB subsystem on the secondary EB subsystem.

{\bf TIC 278352276:} This system was noted by K22 to have ``prominent primary and secondary ETVs''.  Similar to the previous example, HRCam observations of this TIC classify it as an HQND with a 38au upper limit on the size of the Q$^+$EB based on the 824nm filter (Tables \ref{tab:nondetections} and \ref{tab:upperlimit}), and this is consistent with the expectation that the primary and secondary ETVs are being induced by the overall compact nature of the quadruple system, and the gravitational interaction of the two binary pairs with each other.  It's therefore no surprise that we were not able to resolve this source.

{\bf TIC 45160946:} K22 find this system to show prominent ETVs in the primary and secondary, and yet, as seen in Tables \ref{tab:speckle measurements} and \ref{tab:linearsize}, we resolve this source into two components separated by about 90au.  Because this separation is far too great to induce the sizable ETVs found by K22, which show a periodicity of $\sim200$ days, the secondary source we have identified is either an unrelated foreground or background object that just happens to lie very close to the Q$^+$EB, or it could represent part of a higher hierarchy for the system, i.e., the X in a (2 + 2) + X system.  According to Table \ref{tab:speckle measurements}, the HRCam observations found (roughly) a $\Delta m \sim 1.8$ mag between the speckle-resolved components. This magnitude difference is not too far off from that one would expect (i.e., $\Delta m \sim 1.5$ mag) for a (2 + 2) + 1 system of equal luminosity stars. This possibility merits further investigation, e.g., a combination of geometric, photometric, and/or spectroscopic parallaxes to establish whether the various subcomponents are indeed all at the same distance and therefore gravitationally bound.

{\bf TIC 307119043:} We have resolved this source with DSSI (Table \ref{tab:speckle measurements}) to have a relatively wide separation of 0.83 arcseconds. In K22 this system was noted to have ``potential primary ETVs''.  Subsequent investigation using all currently available TESS data bears out the ETV signal for this source;  while the $\sim$$1,500$ day coverage of the TESS data is not sufficient to determine uniquely the periodicity of that signal, it is likely to be on the order of hundreds of days (Kostov et al., in prep.). This would correspond to an outer orbit much smaller than the estimated projected linear separation of $\sim400$ AU (Table \ref{tab:linearsize}) for the two subcomponents we have resolved, which suggests that binary A of TIC 307119043 might be a relatively compact triple system with an unresolved companion.  In this case, this system would be at least a quintuple of architecture (2 + X) + 2.  In this case it is worth noting that the observed $\Delta m$ between the DSSI-resolved subcomponents is 1.75 mag at 880nm but 2.05 mag at 692nm, showing that the stellar compositions of the subcomponents have temperature (and therefore luminosity) variations. TIC 307119043 is a good candidate for follow-up Speckle Imaging During Eclipse (SIDE) observations (Section \ref{sec:SIDEs}) to verify this proposed architecture.

\subsection{Speckle Imaging During Eclipse (SIDE) Observations}
\label{sec:SIDEs}

\subsubsection{Demonstration with TIC 470710327}

For Q$^+$EBs that have been speckle-resolved it is still not possible {\it a priori} to assign particular TESS eclipses to specific speckle-resolved subsystems with confidence. While it is possible that the speckle-resolved sources represent the separate EBs of the Q$^+$EBs, it is also possible that one of the speckle-resolved sources may be an unrelated foreground or background star, or an associated companion to the Q$^+$EBs as part of a higher order architecture --- e.g., [(2+2)+1]. However, time series speckle imaging during eclipses can pinpoint which speckle-resolved component is associated with which set of eclipses. A similar technique was first attempted by \citet{Howell2019} for the Kepler-13AB binary system during an exoplanet transit event to determine which component of the binary the exoplanet orbits; however, the authors state that their results with speckle photometry ``carry too large of an uncertainty to be of
scientific value in this case".

We demonstrate the {\it speckle imaging during eclipse (SIDE)} technique with the ARC 3.5-m + DSSI speckle observations of TIC 470710327 through two primary eclipses of the K22-denoted A subsystem observed on the nights of UT 19 and 20 October 2023. As a check of the K22 ephemerides for this source, we performed simultaneous photometric lightcurve monitoring of TIC 470710327 using the FlareCam imager on the ARCSAT 0.5-m telescope, also located at APO. Over five hundred 18 second exposures on the night of October 19 and over seven hundred 24 second exposures on the night of October 20 were taken  using the Sloan $g$ filter.

Multiple speckle observations were taken on each night to capture TIC 470710327 both in and out of eclipse. Each observation consisted of multiple 1,000 frame sequences. Table~\ref{tab:speckle eclipse} lists the measured $\Delta m$ values in the two DSSI filters for each 1,000 frame sequence individually, along with the average and standard deviation for each observation. Those results where seeing$\times$separation exceeds the canonical 0.6 arcsec$^2$ limit are marked with an asterisk (*). On the night of October 19 (Dates 23.79756 -- 23.79800) in Table~\ref{tab:speckle eclipse}) the observing conditions were highly variable with particularly poor seeing, and therefore the $\Delta m$ values from that night should be considered upper limits.

\begin{figure}
    \centering
    \includegraphics[width=0.45\textwidth]{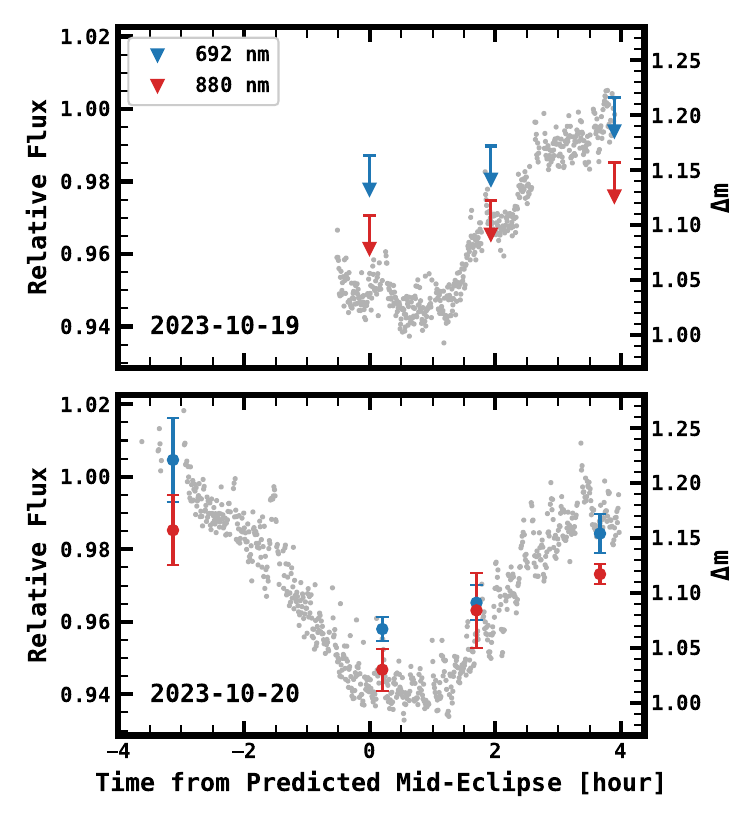}
    \caption{Light curves of TIC470710327 taken on the nights of 19 October (top) and 20 October (bottom) 2023 UT from the ARCSAT telescope at APO. Superposed on the lightcurves are the averaged $\Delta m$ values from DSSI speckle observations using the ARC 3.5-m telescope, with those made in the 692nm filter shown in blue and those made in the 880nm filter shown in red. The averaged $\Delta m$ values on the 19th (top panel) have seeing$\times$separation greater than 0.6, and are therefore considered upper limits. The error bars shown on all $\Delta m$ points are the standard deviations from Table \ref{tab:speckle eclipse}.
    }
    \label{fig:TIC_4707_Speckle-Eclipse}
\end{figure}

Figure \ref{fig:TIC_4707_Speckle-Eclipse} shows the lightcurves from the UT 19-20 October 2023 ARCSAT photometry, which captured two instances of the A subsystem primary eclipse, along with the average speckle imaging $\Delta m$ values in the 692nm and 880nm filters ({\it blue and red symbols}, respectively). In both filters, the general trend is for the $\Delta m$ between the speckle-resolved sources in TIC 470710327 to follow the overall behavior of the lightcurve, such that as the eclipse of subsystem A increases, the $\Delta m$ between the speckle-resolved sources decreases, and as the eclipse goes into egress, the $\Delta m$ increases again. This correspondence demonstrates that subsystem A of TIC 470710327 must lie within the primary speckle-resolved component, i.e., the brighter of the two speckle-resolved sources. This allows us to rule out the possibility that both the A and B subsystems reside within the speckle-resolved companion, or secondary (i.e., the fainter speckle-resolved source). Two possible scenarios therefore remain: either both the A and B subsystems reside within the speckle-resolved primary, or the A and B subsystem reside within the speckle-resolve primary and secondary, respectively. To determine which scenario is correct, SIDE observations during a B subsystem eclipse are required.

\begin{deluxetable*}{lDDlDDlDDlDD}
\tabletypesize{\scriptsize}
\tablewidth{0pt} 
\tablecaption{Speckle Measurements During Eclipse for TIC470710327\label{tab:speckle eclipse}}
\tablehead{
\colhead{Date} & \multicolumn{4}{c}{$\Delta m$ Point \#1} & \colhead{Date} & \multicolumn{4}{c}{$\Delta m$ Point \#2} & \colhead{Date} & \multicolumn{4}{c}{$\Delta m$ Point \#3} & \colhead{Date} & \multicolumn{4}{c}{$\Delta m$ Point \#4} \\
\cline{2-5} \cline{7-10} \cline{12-15} \cline{17-20}
\colhead{(2000+)} & \multicolumn2c{692(nm)} & \multicolumn2c{880(nm)} & \colhead{(2000+)} & \multicolumn2c{692(nm)} & \multicolumn2c{880(nm)} & \colhead{(2000+)} & \multicolumn2c{692(nm)} & \multicolumn2c{880(nm)} & \colhead{(2000+)} & \multicolumn2c{692(nm)} & \multicolumn2c{880(nm)}
}
\decimals
\startdata
23.79756 & 1.12  & 1.11 & 23.79778 & 1.15* & 1.09* & 23.79800 & 1.23  & 1.16 \\
& 1.16* & 1.12* & & 1.22  & 1.16 & & 1.25  & 1.13 \\
& 1.16  & 1.11 & & 1.15  & 1.14 & & 1.22* & 1.21* \\
& 1.17* & 1.10* & & 1.17* & 1.12* & & 1.21* & 1.17* \\
& 1.22* & 1.13* & & 1.17* & 1.10* & & 1.17* & 1.11* \\
& 1.15* & 1.08 & & . & . & & . & . \\
& 1.16* & 1.11* & & . & . & & . & . \\
\cline{2-5} \cline{7-10} \cline{12-15} \cline{17-20}
{\bf Avg.}& \bf1.\bf163* & \bf1.\bf109* & & \bf1.\bf172* & \bf1.\bf122* & & \bf1.\bf216* & \bf1.\bf156* \\
{\bf Stdev} & \bf0.\bf030 & \bf0.\bf016 & & \bf0.\bf029 & \bf0.\bf029 & & \bf0.\bf030 & \bf0.\bf038 \\
\hline
23.80023 & 1.20  & 1.21 & 23.80061 & 1.06  & 1.05 & 23.80078 & 1.09  & 1.06 & 23.80100 & 1.15  & 1.10 \\
& 1.29  & 1.16 & & 1.08  & 1.05 & & 1.07  & 1.07 & & $\cdots$ & 1.12 \\
& 1.24  & 1.13 & & 1.06  & 1.00 & & 1.11  & 1.15 & & 1.16  & 1.12 \\
& 1.20  & 1.13 & & 1.08  & $\cdots$ & & 1.10  & 1.07 & & 1.13  & 1.11 \\
& 1.24  & 1.13 & & 1.07  & 1.02 & & 1.11  & 1.05 & & 1.15  & 1.12 \\
& 1.20  & 1.19 & & 1.05  & 1.03 & & 1.08  & 1.10 & & 1.13  & 1.12 \\
& 1.18  & 1.15 & & 1.07  & 1.03 & & 1.08  & 1.09 & & 1.17  & 1.11 \\
& . & . & & . & . & & . & . & & 1.16  & 1.13 \\
& . & . & & . & . & & . & . & & 1.18  & 1.12 \\
\cline{2-5} \cline{7-10} \cline{12-15} \cline{17-20}
{\bf Avg.} & \bf1.\bf22 & \bf1.\bf16 & & \bf1.\bf07 & \bf1.\bf03 & & \bf1.\bf09 & \bf1.\bf08 & & \bf1.\bf15 & \bf1.\bf12 \\
{\bf Stdev} & \bf0.\bf038 & \bf0.\bf032 & & \bf0.\bf011 & \bf0.\bf019 & & \bf0.\bf016 & \bf0.\bf034 & & \bf0.\bf018 & \bf0.\bf009 \\
\enddata
\tablecomments{$\Delta m$ values marked with an asterisk (*) are considered an upper limit because the observation may be affected by speckle decorrelation as discussed in \citet{Horch2004}.}
\end{deluxetable*}

\subsubsection{Serendipitous Speckle Imaging Observations During Eclipses}

Given the number of speckle observations that have been collected and the relatively short periods of many of the binaries within the Q$^+$EBs, one might wonder whether we may have, by chance, imaged any of the other systems while they were experiencing an eclipse.  Such events could partly explain discrepancies in $\Delta m$ that are seen from epoch to epoch for different systems reported in Table \ref{tab:speckle measurements} and contribute to the scatter in $\Delta m$ values for each Q$^+$EB.
To ascertain this possibility, we assumed that the K22 ephemerides still hold and advanced them for each Q$^+$EB system to each time that we observed it to check whether the system would have been experiencing an eclipse at that point.

Aside from the SIDE observations of TIC 470710327, we estimate that 10 of the remaining 34 DSSI observations in Table \ref{tab:speckle measurements} were taken during an eclipse, where the latter is defined as the time from ingress to egress. This is a very general statement, as it depends on the various assumptions made. And while this would seem to be about a third of the observations, whether an impact on $\Delta m$ would be perceptible is highly dependent on the overall eclipse depth and timing phase.  Note also that the converse is not reliably dependable --- that is, variations in $\Delta m$ from  run to run can just as easily reflect systematic variations in observing conditions and data quality and therefore may not be indicative of an eclipse situation.  Discerning these differences requires careful and thorough investigation or, better, controlled, systematic observations like those performed here for TIC 470710327.

\section{Conclusions}

We have begun follow-up observations and investigations of the sample of quadruple and higher order multiplicity eclipsing binaries being discovered within the TESS database.  Here we have focused on multi-filter speckle imaging on 4-m class telescopes of a sample of 57 systems, out of the 97 systems presented by \citet{Kostov2022}. Of the 57 systems investigated here, we are able to resolve 34 (nearly 60\%) into two distinct sources. Using \cite{Bailer2021} geometric distances, we have calculated estimated linear separations between the resolved companions, and have placed upper limits on the linear size for the unresolved systems. Among the former are several  quadruple systems apparently with a {\it projected} linear separation the size of the Solar system ---  TIC 97356407 (32 au), TIC 146810480 (35 au), TIC 328181241 (38 au), and TIC 399492452 (30 au).

We note that the \citet{Bailer2021} distances for the two speckle-resolved components in TIC 266771301 are vastly different, which could mean that the system may be two unrelated EBs along the same line-of-sight, or it is a truly bound \added{Q$^+$EB} residing in  either the $\sim3$ kpc or the $\sim 0.2$ kpc source.  If the latter, the upper limit of the projected angular size of the system would be 10.8 au, the smallest among all of our measured systems. With DSSI observations we also confirm the previously resolved \citep{Powell2021} two-source parameters (separation and position angle) for the sextuple system TIC 168789840, although we obtain differing photometric contrasts between the two sources compared to those seen by those authors.  

Within the systems studied here are four --- two that we have resolved into two sources and two HQNDs --- that have previously been reported to have eclipse timing variations, a sign that an EB is gravitational perturbed by the influence of a nearby star or stars (perhaps the other EB in the \added{Q$^+$EB}).  The two HQNDs are found to have upper limits to the projected system size that are quite compact --- 41au for TIC 52856877 and 38au for TIC 278352276 --- a result consistent with the ETV behavior.  In contrast, the two subcomponents we have resolved for TIC 45160946 are found to have a projected separation of $\sim 90$au --- far too large for this to explain the 200 day periodicity in the ETV --- it is likely that the speckle-resolved secondary is either a fore/background contaminant unrelated to the \added{Q$^+$EB}, which must be entirely bound up within the primary, or that the secondary represents part of additional hierarchy in the system outside the (2 + 2) \added{Q$^+$EB} represented by the primary.  Meanwhile, we argue that the wide, 0.83 arcsec  separation of the two resolved subcomponents of TIC 307119043 combined with the likely several hundred day period of the ETV is evidence that the system is likely at least a quintuple of (2 + X) + 2 architecture.

Finally, we have also for the first time successfully demonstrated the technique of {\it speckle imaging during eclipse} (SIDE) as a method to ascertain which of the speckle-resolved components is associated with a particular EB eclipse family, and a new tool for placing further constraints on the possible configurations of \added{Q$^+$EB} systems.

\added{In future contributions we will present results from our continuing campaign of speckle imaging for other systems from \cite{Kostov2022}, as well as those from other catalogs of TESS Q$^+$EBs, using 4-m class telescopes, as reported here.  In addition, we will also report on speckle imaging conducted with 8-m class telescopes for targets too faint to observe with 4-m class telescopes, and for Q$^+$EB systems that were unresolved by speckle imaging with 4-m class telescopes.  Finally, we will present results of the application of the SIDE technique for systems we have previously resolved.}


\begin{acknowledgments}

We are grateful to the very helpful APO staff for accommodating DSSI as a visiting instrument on the ARC 3.5-m telescope. S.R.M., J.W.D., E.F., G.H., and V.K. are also grateful for funding from NASA ADAP award 80NSSC21K0631 (V.K., Principal Investigator) to the SETI Institute, including a subaward to the University of Virginia for the UVa participants. J.W.D. gratefully acknowledges funding from the NASA-NSF Exoplanet Observational Research (NN-EXPLORE) program through JPL RSAs 1560091 and 1569072, as well as funding from the RECONS Institute. E.P.H. acknowledges funding from NSF grants AST-1909560 and AST-2206099. S.R.M., J.W.D., and E.F. also acknowledge support from The Research Corporation via a Cottrell Singular Exceptional Endeavors of Discovery (SEED) Award. The material is based upon work supported by NASA under award number 80GSFC24M0006.

This research has made use of the SIMBAD database, operated at CDS, Strasbourg, France, as well as the Exoplanet Follow-up Observation Program (ExoFOP; \citealt{NExScI-ExoFOP}) website, which is operated by the California Institute of Technology, under contract with the National Aeronautics and Space Administration under the Exoplanet Exploration Program.

This work presents results from the European Space Agency (ESA) space mission Gaia. Gaia data are being processed by the Gaia Data Processing and Analysis Consortium (DPAC). Funding for the DPAC is provided by national institutions, in particular the institutions participating in the Gaia MultiLateral Agreement (MLA). The Gaia mission website is \url{https://www.cosmos.esa.int/gaia}. The Gaia archive website is \url{https://archives.esac.esa.int/gaia}.

This paper includes data collected with the TESS mission, obtained from the MAST data archive at the Space Telescope Science Institute (STScI). Funding for the TESS mission is provided by the NASA Explorer Program. STScI is operated by the Association of Universities for Research in Astronomy, Inc., under NASA contract NAS 5–26555.

\end{acknowledgments}

\vspace{5mm}
\facilities{APO(Visiting Instrument DSSI), APO:ARCSAT(FlareCam), SOAR(HRCam)}

\software{astropy \citep{Astropy2013,Astropy2018},  
          AstroImageJ \citep{Collins2017}, numpy \citep{numpy}, matplotlib \citep{matplotlib}, pandas \citep{pandasarticle,pandassoftware} 
          }

\bibliography{Quad-paper}{}
\bibliographystyle{aasjournalv7}

\end{document}